\documentclass[
    aip,
    pop,
    amsmath,
    amssymb,
    reprint,
    floatfix
]{revtex4-1}
\usepackage{graphicx}
\usepackage{float}
\usepackage{dcolumn}
\usepackage[mathlines]{lineno}

\usepackage[utf8]{inputenc}
\usepackage[T1]{fontenc}
\usepackage{mathptmx}
\usepackage{etoolbox}

\draft{}

\usepackage{amssymb}
\usepackage{amsfonts}
\usepackage{graphicx}
\usepackage{subcaption}
\usepackage{multirow}
\usepackage{bigstrut}
\usepackage{epstopdf}
\usepackage{upgreek}
\usepackage[english]{babel}
\usepackage{csquotes}
\usepackage{combelow}
\usepackage{amsmath}
\usepackage{amsthm}
\usepackage{xcolor}
\usepackage{bm}
\usepackage{tikz}
\usetikzlibrary{shapes}
\usetikzlibrary{arrows}
\usetikzlibrary{positioning}
\usetikzlibrary{matrix}
\usetikzlibrary{patterns}
\usetikzlibrary{decorations.pathreplacing,decorations.pathmorphing,decorations.markings}
\usepackage{tikz-qtree}
\usetikzlibrary{trees,calc,arrows.meta,positioning,bending}
\usepackage{mathtools}
\usepackage{hyperref}

\hypersetup{
    pdftitle={Machine Learning for Electron-Scale Turbulence Modeling in W7-X},
    pdfauthor={Ionut-Gabriel Farcas, Don Lawrence Carl Agapito Fernando, Alejandro Ba\~n\'on Navarro, Gabriele Merlo, Frank Jenko},
    pdfkeywords={machine learning surrogate, surrogate modeling stellerators, W7-X, ETG transport, ETG transport stellarators, reduced model ETG, reduced model ETG W7-X, machine learning surrogate W7-X, machine learning stellarators}
}

\definecolor{carminepink}{rgb}{0.92, 0.3, 0.26}
\definecolor{cobalt}{rgb}{0.0, 0.28, 0.67}
\definecolor{cerulean}{rgb}{0.0, 0.48, 0.65}

\newcount\Comments
\newcommand{\kibitz}[2]{\ifnum\Comments=1\textcolor{#1}{#2}\fi}

\begin{document}

\title{Machine Learning for Electron-Scale Turbulence Modeling in W7-X}

\author{Ionu\c{t}-Gabriel Farca\c{s}}
\email{farcasi@vt.edu}
\affiliation{Department of Mathematics and Division of Computational Modeling and Data Analytics, Academy of Data Science, Virginia Tech, Blacksburg, Virginia 24061}
\altaffiliation{I.~Farca\c{s} and D.L.C.~Agapito Fernando contributed equally to this work}

\author{Don Lawrence Carl Agapito Fernando}
\email{don.fernando@ipp.mpg.de}
\affiliation{Max Planck Institute for Plasma Physics, 85748 Garching, Germany}
\altaffiliation{I.~Farca\c{s} and D.L.C.~Agapito Fernando contributed equally to this work}

\author{Alejandro Ba\~n\'on Navarro}
\email{alejandro.banon.navarro@ipp.mpg.de}
\affiliation{Max Planck Institute for Plasma Physics, 85748 Garching, Germany}

\author{Gabriele Merlo}
\email{gabriele.merlo@ipp.mpg.de}
\affiliation{Max Planck Institute for Plasma Physics, 85748 Garching, Germany}

\author{Frank Jenko}
\email{frank.jenko@ipp.mpg.de}
\affiliation{Max Planck Institute for Plasma Physics, 85748 Garching, Germany}

\begin{abstract}
Constructing reduced models for turbulent transport is essential for accelerating profile predictions and enabling many-query tasks such as parameter exploration, uncertainty quantification, and design optimization. 
This work investigates machine-learning-driven reduced models for Electron Temperature Gradient (ETG) turbulence in the Wendelstein 7-X (W7-X) stellarator. 
We develop physics-guided scaling laws to predict the ETG heat flux at seven radial locations as functions of three key plasma parameters: the normalized electron temperature gradient ($\omega_{T_e}$), the ratio of normalized electron temperature and density gradients ($\eta_e$), and the electron-to-ion temperature ratio ($\uptau$). 
The model coefficients are determined through regression combined with an active learning strategy. 
The procedure initializes the scaling laws using low-cardinality sparse-grid training data and iteratively enriches the training set by selecting maximally informative samples from an existing simulation database. 
The predictive performance of the models is assessed using out-of-sample datasets comprising more than $393$ points per radial location. 
Using the coefficients identified at the seven training radial locations, we further derive regression-based parameterizations for the scaling-law coefficients as functions of radial position. 
The resulting models are then evaluated at three additional radial locations not used during training, including both interpolation and moderate extrapolation cases.
Overall, our reduced models demonstrate good predictive performance and achieve accuracy comparable to the original reference simulations, including in interpolation and moderate extrapolation regimes. 
An important finding is that a single radius-independent model cannot adequately describe ETG transport across the W7-X core, suggesting the presence of geometry-dependent physics not captured by the present formulation.
\end{abstract}


\maketitle 

\section{Introduction}\label{sec:intro}
The scientific landscape is being reshaped by rapid advances in data-driven modeling and machine learning (ML). 
One field poised to benefit significantly is the study of turbulent transport in magnetic confinement fusion devices -- a central challenge in fusion research. 
Understanding plasma turbulence is a key prerequisite for designing optimized fusion power plants. 
Although high-fidelity first-principles simulations are essential for uncovering the complex physics involved, their high computational cost limits their routine use for rapid profile predictions and many-query tasks, such as uncertainty quantification, parameter scans, and design optimization, which usually require large ensembles of expensive simulations. 
These applications demand reduced models with high computational efficiency and predictive reliability.

This paper investigates the efficient construction of robust reduced models for electron temperature gradient (ETG) turbulence in the Wendelstein 7-X (W7-X) stellarator.
For over two decades, ETG turbulence has been a significant subject of theoretical and numerical investigation. 
Early seminal works~\cite{Je00, Dorland, JD02} primarily focused on tokamak core plasmas, whereas more recent studies have increasingly indicated its role in regulating transport within the tokamak pedestal region~\cite{told, jenko_09, Idomura, Hatch_2015, Hatch_2017, Hassan_2022, Chapman-Oplopoiou_2022, Li_2024, Pa20, Pa22, St22, Le23, BCS23, BCS24} and in stellarators~\cite{Jenko_kendl_2002, Plunk_PRL_2019, Weir_2021}. 
Recognizing this importance, several recent studies~\citep{Chapman-Oplopoiou_2022, Guttenfelder_2021, Ha22, Hatch_2024,FMJ24} developed reduced models to describe ETG heat fluxes in the tokamak pedestal. 
For instance, \citet{Ha22} derived simple algebraic formulae for the electron heat flux using a database of $61$ discharges from JET, DIII-D, ASDEX Upgrade, and C-MOD experiments. 
These formulae were further developed and extended to scaling laws by~\citet{Hatch_2024} using a refined version of this database. 
In a separate effort, \citet{FMJ24} formulated a reduced model as a scaling law using structure-exploiting sparse grids~\cite{FMJ22} and evaluated its prediction capabilities against within-distribution, out-of-bounds, and out-of-distribution datasets. 
Other recent work by~\citet{Turica_Field_Schekochihin_Frassinetti_2025} employed turbulence models and ML to predict electron temperature profiles in the JET pedestal region.

While progress has been made in reduced modeling for ETG transport in tokamaks, reduced ETG transport models for stellarators remain largely underexplored.
This work aims to address this gap by constructing reduced ETG turbulence models for W7-X.
We investigate the construction of physics-guided scaling laws comprising three key plasma  parameters: the normalized electron temperature radial gradient ($\omega_{T_e}$), the ratio  of normalized electron temperature and density radial gradients ($\eta_e$), and the electron-to-ion temperature ratio ($\uptau$).
We employ an ML-driven strategy that combines regression and active learning to efficiently fit the scaling law coefficients from small yet informative datasets.
Our process begins by constructing an initial set of models from datasets of $11$--$12$ points generated using a structure-exploiting sparse grid approach~\cite{FMJ22}.
These models are then iteratively refined through an active-learning procedure that selects the most informative training data pairs.
This procedure is driven by computing a measure of prediction uncertainty at each step over a test set of candidate input parameter triplets $(\omega_{T_e}, \eta_e, \uptau)$.
To quantify this uncertainty, we use bootstrapping~\cite{EB86} to compute $95\%$ prediction intervals for the ETG heat flux, along with the mean prediction, for all triplets in the test set.
A prediction interval defines a range that, with a specified probability (e.g., $95\%$), 
is expected to contain a future individual observation, conditional on the assumed model 
form.
Bootstrapping provides a robust, distribution-free method for estimating such intervals without requiring specific assumptions about the underlying data or noise distributions.
We select the input triplet from the test set for which the percentage deviation of the prediction interval from the mean prediction is largest, a criterion that is independent of the reference simulated heat flux value.
The selected triplet, together with its corresponding heat flux computed from a high-fidelity simulation, is then appended to the training dataset, and the model is retrained on the augmented data.
To prevent premature termination, we additionally compute the relative error between the newly added heat flux value and the prediction obtained at the previous active learning step.
The procedure terminates when both the maximum percentage deviation of the prediction interval and this relative error fall below user-prescribed thresholds.
Active learning has previously been applied to construct reduced models for fusion plasmas (e.g., Refs.~\citenum{Ro22, Ho25}); however, to our knowledge, its application in the 
context of stellarators remains unexplored.

Active learning is typically implemented in an online setting, where new heat flux values are computed after the most informative input parameter values are selected by the algorithm.
In the present paper, however, we leverage a pre-existing database of ETG simulations generated in the context of the validation study for the \texttt{\textsc{Gene}-KNOSOS-Tango} framework \cite{Fernando2025}. 
A subset of the input parameter values from this database is treated as the active learning testing set.
Consequently, the procedure identifies the best model, according to the metrics guiding the active learning process, that can be supported by the available database.
The database was generated by the application of the integrated \texttt{\textsc{Gene}-KNOSOS-Tango} simulation framework, which self-consistently couples various physics codes to predict steady-state plasma profiles, to four experimental OP1.2b W7-X scenarios.~\cite{Fernando2025} 
Our work contributes to a growing interest in using data-driven techniques to build reduced turbulence models for stellarators.
For example, a recent study by~\citet{landreman2025doesiontemperaturegradient} investigated using multiple ML techniques including convolutional neural networks to develop surrogate models of ion-temperature-gradient turbulent transport from a database of $200,000$ adiabatic-electron gyrokinetic runs.

We apply this ML-based framework to construct reduced models at seven radial locations across W7-X, requiring at most $190$ training samples per location. 
Generalization performance is assessed using the remaining portion of the pre-existing database, comprising more than $393$ out-of-sample points per radial location, with predictive uncertainty quantified through $95\%$ bootstrapped prediction intervals. 
We furthermore investigate the extension of the reduced models beyond the original seven radial locations by regressing the inferred scaling-law coefficients against radial position. 
The resulting regression models provide explicit functional dependencies for the coefficients and are used to construct reduced models at three additional radial locations, including both interpolation and moderate extrapolation cases. 
The resulting models exhibit good predictive accuracy across all radial locations considered.  
To our knowledge, this work represents the first systematic effort to derive reduced ETG transport models across multiple radial locations in a stellarator.

The remainder of this paper is organized as follows. 
Section~\ref{sec:target_models} introduces the target reduced models formulated as scaling laws. 
Section~\ref{sec:datasets} describes the datasets used for model development and numerical validation. 
Our ML-driven approach for constructing these models for ETG transport in W7-X at seven radial locations is presented in Sec.~\ref{sec:ML_based_reduced_modeling}. 
In Sec.~\ref{sec:results_original_seven_loc}, we assess the models' generalization performance on out-of-sample data. 
Section~\ref{sec:results_three_further_loc} extends this analysis to arbitrary radial locations via regression of the coefficients fitted at the original seven positions. 
Section~\ref{sec:conclusions} concludes the paper.

\section{Target reduced models for ETG turbulence in W7-X} \label{sec:target_models}

We target reduced models for ETG turbulence in W7-X as scaling laws based on the following ansatz:
\begin{equation} \label{eq:target_model}
Q_e/Q_{\mathrm{GB}}[\hat{\rho}] = c_0(\hat{\rho}) \, \omega_{T_e}^{p_1(\hat{\rho})} \left(\eta_e - 1\right)^{p_2(\hat{\rho})} \uptau^{p_3(\hat{\rho})}\,
\end{equation}
where $\hat{\rho} = \rho/a$ denotes the radial location, $Q_e$ is the electron heat flux (in $\mathrm{W/m^{2}}$), and $Q_{\mathrm{GB}} = n_e T_e^{5/2} m_e^{1/2} / (e B_{\rm ref} a)^2$ is the flux in gyro-Bohm (GB) units.
Here, $B_{\rm ref}$ denotes the magnetic field strength on axis (in T), $n_e$ is the electron density (in $10^{19}~\mathrm{m}^{-3}$), $T_e$ is the electron temperature (in keV), $m_e$ the electron mass, and $a$ is the effective minor radius of the magnetic geometry, defined as $a = \sqrt{{\Psi_{\rm LCFS}}/{\pi B_{\rm ref}}}$.
The quantity $\Psi_{\rm LCFS}$ denotes the toroidal magnetic flux through the last closed flux surface (LCFS).
The dimensionless radial coordinate $\hat{\rho} = \sqrt{{\Psi}/{\Psi_{\rm LCFS}}}$ is defined by the ratio between the $\Psi$ value of a flux surface and that of the LCFS.
This model depends on four coefficients $\{c_0(\hat{\rho}), p_1(\hat{\rho}), p_2(\hat{\rho}), p_3(\hat{\rho})\} \in \mathbb{R}^4$.

The scaling law~\eqref{eq:target_model} is a generic expression that captures the known dependencies of ETG turbulence on its driving parameters.
It relates $Q_e/Q_{\mathrm{GB}}$ to three key parameters: the normalized electron temperature radial gradient, $\omega_{T_e} = -(a/T_e)\;(dT_{e}/d\rho)$, the ratio of normalized electron temperature and density radial gradients, $\eta_e = \omega_{T_e}/\omega_{n_e}$, and the electron-to-ion temperature ratio, $\uptau = Z_{\rm eff}T_e/T_i$, where $T_i$ denotes the ion temperature.
$Z_\mathrm{eff}$ is the effective ion charge retained in the collision operator, which in our simulations is the linearized Landau--Boltzmann operator.
Our simulations use hydrogen ions and do not account for impurities, leading to $Z_\mathrm{eff} = 1$.
The selection of these three input parameters is partially motivated by previous studies~\cite{Je00,Jenko_pop_2001,Hatch_2024,FMJ24}, as well as by the expectation that ETG turbulence in the W7-X core is predominantly slab-like due to the combination of low magnetic shear and short connection lengths~\cite{Plunk_PRL_2019, Merlo2026}.
We further note that the scaling law in Eq.~\eqref{eq:target_model} is similar to those considered in~\citet{Hatch_2024} and~\citet{FMJ24} in the context of reduced models for ETG turbulence in the tokamak pedestal.

A few compromises and simplifications were introduced in the functional form~\eqref{eq:target_model}.
First, the $\eta_e$ threshold is not expected to be exactly unity in W7-X and may vary with radial position.
Preliminary curve fittings confirm this, yielding threshold values that vary with $\hat{\rho}$ and, consequently, with the local geometry.
However, a complementary set of electron-scale simulations scanning $\eta_e$ has shown that the threshold is close to unity for a slab-like ETG branch~\cite{Jenko_pop_2001,
Jenko_kendl_2002,Guttenfelder_2021}, suggesting that adopting a fixed threshold of unity is a reasonable and practical approximation for the present models.
This choice also mitigates convergence issues that can arise in the active-learning-driven regression procedure when the $\eta_e$ threshold is treated as a free model coefficient.
As further simplifications, cross-couplings between input parameters and dependencies of the $\eta_e$ threshold on radial position are not incorporated into the present model.
Extensions of the scaling law along these lines are left for future work, to be carried out with purpose-built training datasets.

We first construct scaling laws~\eqref{eq:target_model} at seven radial locations, 
$\hat{\rho} \in \{0.2, 0.3, 0.4, 0.5, 0.6, 0.7, 0.8\}$ (Sec.~\ref{sec:ML_based_reduced_modeling}). 
Subsequently, we develop generalized models via coefficient regression to enable predictions beyond these locations. 
The predictive performance of the resulting models is evaluated at three additional radial positions, $\hat{\rho} \in \{0.1, 0.55, 0.75\}$ (Sec.~\ref{sec:results_three_further_loc}).
The use of radius-specific models rather than a single generalized model reflects the difficulty of representing the local magnetic geometry in a non-axisymmetric device using only a small set of parameters, as is often feasible in the core region of a tokamak.
For completeness, we nevertheless also explored the construction of a generic model based on the form of Eq.~\ref{eq:target_model}, but with radially independent coefficients. 
A discussion of these results is provided in Appendix~\ref{sec_app:generic_model}.

\section{Datasets for reduced model development and numerical validation}\label{sec:datasets}

This section describes the datasets used to construct and numerically validate our reduced models for ETG transport in W7-X.
Section~\ref{subsec:SG_initmodel} presents the low-cardinality sparse grid datasets used to build the initial models across seven radial positions. 
Section~\ref{subsec:hi_fi_database} details the pre-existing database of high-fidelity simulations at the same positions.
A fraction of this database is used for iterative model refinement while the remainder is employed for assessing the models' generalization capabilities.
Finally, Sec.~\ref{subsec:database_three_additional_loc} presents the data for three additional radial positions used to investigate the coefficient regression-based models.

\subsection{Low-cardinality sparse grid datasets for constructing the initial reduced models across seven radial locations}
\label{subsec:SG_initmodel}
Before performing active learning, we require initial (base) models for each $\hat{\rho} \in \{0.2, 0.3, 0.4, 0.5, 0.6, 0.7, 0.8\}$, corresponding to the radial locations simulated in the validation study~\cite{Fernando2025} of the \texttt{\textsc{Gene}-KNOSOS-Tango} framework discussed in the following section. 
Initial training datasets for these locations are generated using a structure-exploiting sparse grid method~\cite{FMJ22}, previously shown to be effective for constructing predictive reduced models of ETG transport in the tokamak pedestal~\cite{FMJ24}. 
This method exploits anisotropy in the input parameter space, enabling efficient sampling by concentrating adaptive refinement on the most influential parameters and their interactions.
Although the sparse grid method could, in principle, be used directly to construct the target ETG reduced models, doing so would require careful selection of parameter bounds for each $\hat{\rho}$. 
In the present work, however, we instead leverage a pre-existing high-fidelity simulation database (Sec.~\ref{subsec:hi_fi_database}). 
Accordingly, the sparse grid approach is used only to construct initial reduced models with a small number of adaptive refinement steps for each $\hat{\rho}$. 
These initial models are then iteratively improved through active learning by selecting the most informative samples from the high-fidelity database. 
A more detailed discussion and justification for using sparse-grid-generated data to initialize the active learning procedure is provided in Appendix~\ref{sec_app:active_learning_sensitivity}.

In the simulations used to obtain the sparse grid datasets, the plasma behavior is modeled using six local inputs, $\{\upbeta_e, \uplambda_{\mathrm{De}}, \upnu_c, \uptau, \omega_{n_e}, \omega_{T_e}\}$, where $\upbeta_e=403\times10^{-5} n_e T_e/B^{2}_{\rm ref}$ denotes the ratio of plasma pressure to magnetic pressure and $\uplambda_{De} = 0.9859 \times B_{\rm ref} n^{-1/2}_{\rm e}$ is the normalized Debye length.
The normalized collision frequency is given by $\nu_c = 2.3031 \times 10^{-5} a(\ln{\Lambda})  n_{\rm e} T^{-2}_{\rm e}$, where $\ln{\Lambda}$ is the Coulomb logarithm.
The specific bounds for the six local plasma parameters are summarized in Table~\ref{tab:SG_parameters}. 
As in Ref.~\citenum{FMJ24}, we assume a uniform distribution within these bounds, which are held constant across all $\hat{\rho}$.
In this setup, $\eta_e$ varies between $1.5$ and $50.0$.
\begin{table}[tb]
\centering
\begin{tabular}{c|c|c}
\hline
\hline
Input parameter & Lower bound & Upper bound \\
\hline
$\upbeta_e$ & $ 4.00 \times 10^{-4}$ & $ 5.00 \times 10^{-2}$\\
$\uplambda_{\mathrm{De}}$ & $0.0$ & $1.73$ \\
$\upnu_c$& $1.00 \times 10^{-5}$ & $1.00 \times 10^{-3}$ \\
$\uptau$& $0.80$ & $2.50$ \\
$\omega_{n_e}$ &  $0.10$ & $1.00$\\
$\omega_{T_e}$ & $1.50$ & $5.00$ \\
\hline
\hline
\end{tabular}
\caption{Summary of the six local plasma parameters and their uniform bounds. These bounds define the input space for the sparse grid approach used to generate the low-cardinality datasets for the initial reduced models.}
\label{tab:SG_parameters}
\end{table}

These adiabatic-ion simulations were performed using the plasma micro-turbulence gyrokinetic simulation code \textsc{Gene}~\citep{Je00} in the flux-tube limit using a box with $n_{k_x}\times n_{k_y}\times n_{z}\times n_{v_\|}\times n_{\mu}=192\times72\times96\times32\times8$ degrees of freedom in the five-dimensional position--velocity space.
Key simulation parameters include $k_{y,min}\,\rho_e=0.025 $ and $L_x\simeq 350\,\rho_e$, where $\rho_e=v_e/\Omega_e$ is the electron gyroradius, $v_e=\sqrt{T_e/m_e} = v_{th}/\sqrt{2}$ is the electron reference velocity, and $\Omega_e = e B_{\rm ref}/m_e$. 
In velocity space, a box characterized by $-3\leq v_\|/v_{th}\leq 3$ and $0\leq\mu B_{\rm ref} /  T_e\leq9$ has been used. 
We conducted several convergence tests, including runs at the extrema of the two gradient parameters, to confirm that the employed \textsc{Gene} grid resolution was sufficient for accurate results.

Table~\ref{tab:SG_database_parameters} presents the number of adaptive refinement steps, the cardinality of the corresponding sparse grid, denoted by $N_{\mathrm{SG}} \in \mathbb{N}$, and the minimum and maximum values of the computed heat flux $Q_e/Q_{\mathrm{GB}}$ for all $\hat{\rho}$. 
The adaptive refinement procedure was intentionally limited to four to five steps, yielding low-cardinality grids containing only $11$--$12$ points. 
This relatively small number of refinement steps was chosen to construct lightweight base models for the subsequent active learning procedure. 
The base models are then systematically refined using data from the pre-existing database.
\begin{table}[tb]
\centering
\begin{tabular}{c|c|c|c|c}
\hline
\hline
$\hat{\rho}$ & Refinement steps & $N_{\mathrm{SG}}$ & $Q_{e, \mathrm{min}}/Q_{\mathrm{GB}}$ & $Q_{e, \mathrm{max}}/Q_{\mathrm{GB}}$ \\
\hline
$0.2$ & $4$ & $11$ & $0.73$ & $52.6$\\
$0.3$ & $5$ & $12$ & $0.91$ & $63.1$\\
$0.4$ & $4$ & $11$ & $0.83$ & $69.5$\\
$0.5$ & $4$ & $11$ & $1.10$ & $73.2$\\
$0.6$ & $4$ & $11$ & $0.96$ & $78.8$\\
$0.7$ & $5$ & $12$ & $0.09$ & $87.4$\\
$0.8$ & $5$ & $12$ & $0.11$ & $97.6$\\
\hline
\hline
\end{tabular}
\caption{Number of refinement steps, sparse grid cardinality, and minimum/maximum heat flux values obtained using the initial sparse grid approach.}
\label{tab:SG_database_parameters}
\end{table}

\subsection{Database for iterative model refinement and numerical validation at seven radial locations}\label{subsec:hi_fi_database}
\begin{figure}[htb]
    \centering
    \includegraphics[width=0.40\textwidth]{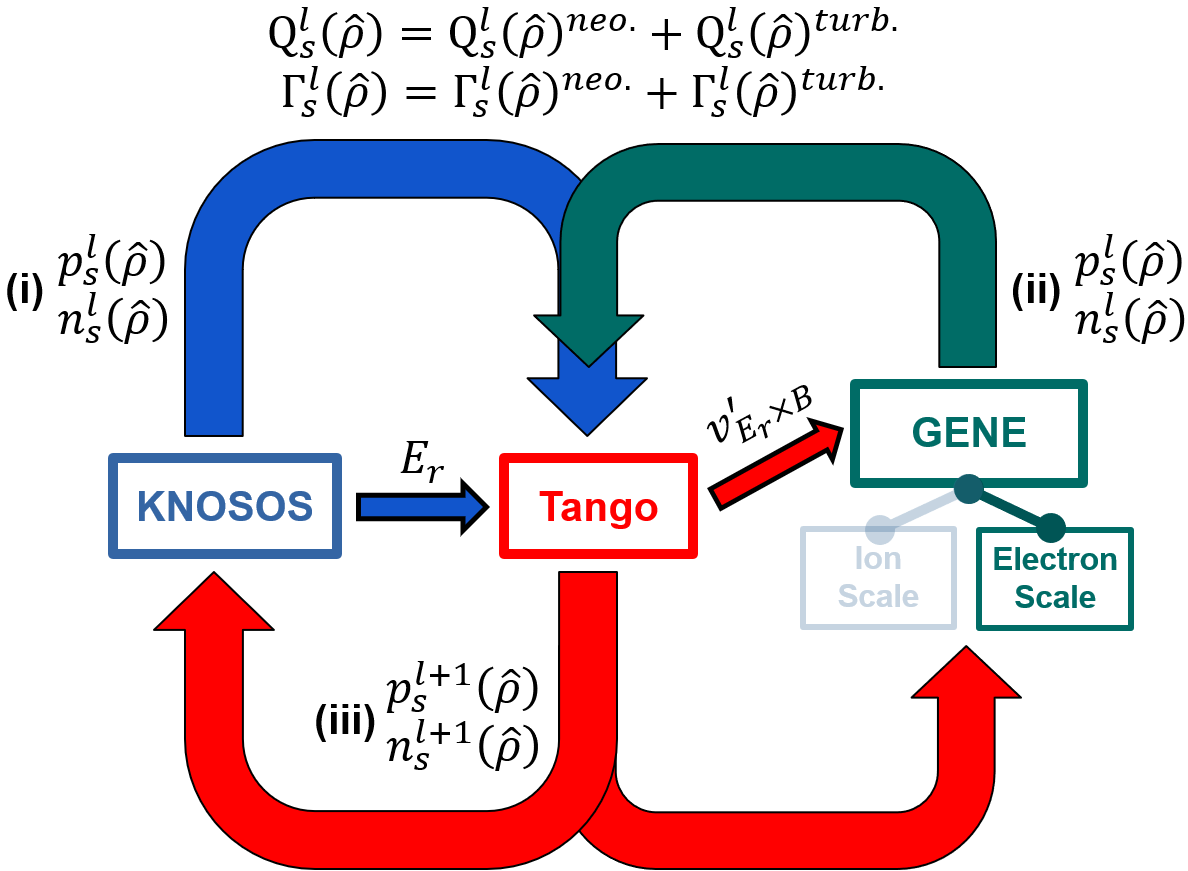}
    \caption{\label{fig:GTK_framework}
        The \texttt{\textsc{Gene}-KNOSOS-Tango} framework evolves the plasma density ($n_e$) and pressure ($p_e,\;p_i$) profiles. Plasma profiles at iteration $l$ are provided to (i) \texttt{KNOSOS} and (ii) \textsc{Gene} to compute the neoclassical and turbulent fluxes, respectively. These are then summed and passed to (iii) \texttt{Tango}, which generates updated plasma profiles for iteration $(l + 1)$. The iteration loop continues until satisfactory agreement is reached between the fluxes and their corresponding sources (energy and/or particles). A simulation database is constructed from the input parameters, profiles, and fluxes obtained at each iteration. Since only the electron-scale adiabatic-ion simulations from the database are used in this study, the ion-scale simulations from \textsc{Gene} have been grayed out.}
\end{figure} 

The high-fidelity data for active-learning-based iterative model refinement and numerical validation were generated using the \texttt{\textsc{Gene}-KNOSOS-Tango} simulation framework. 
This framework couples \textsc{Gene}, the neoclassical transport code \texttt{KNOSOS}~\cite{Velasco2020,Velasco2021}, and the one-dimensional transport solver \texttt{Tango}~\cite{Shestakov2003,Parker2018} to predict the steady-state plasma temperature and density profiles in stellarators.
The workflow is summarized in Fig.~\ref{fig:GTK_framework}.
By comparing the heat and particle fluxes to their respective sources at each iteration, \texttt{Tango} generates new plasma profiles for each successive \textsc{Gene} and \texttt{KNOSOS} simulation. 
The variations in the plasma profiles depend on how well the heat and particle balances are satisfied. 
As a rule of thumb, fluxes that overshoot the sources at any position lead to a local flattening of the profile while fluxes that underestimate the sources result in a steepening instead. 

The database used in the present paper consists of the ensemble of all steady-state adiabatic-ion flux-tube simulations used by the \texttt{\textsc{Gene}-KNOSOS-Tango} framework~\cite{Fernando2025} to obtain steady-state temperature and density profiles for four W7-X scenarios, namely the ``low-density ECRH", ``high-density ECRH", ``NBI + ECRH", and ``NBI" cases stemming from three OP1.2b experimental campaign discharges.~\cite{Carralero2021,Carralero2022}
Simulations where the electron-scale heat fluxes displayed transient time behavior are excluded from the database.
The naming convention reflects the type of heating in these scenarios, which are electron cyclotron resonance heating (ECRH) and neutral-beam injection (NBI).
\textsc{Gene} simulations were performed at several radial positions, including those considered in this paper, with the standard magnetic configuration of W7-X while exclusively using the field line passing through the outboard mid-plane of W7-X's bean-shaped cross-section.
The target database consists of the electron-scale turbulent heat fluxes $Q_{\rm e}/Q_{\rm GB}$ and their corresponding input parameters $\{\omega_{T_e}, \eta_e, \uptau\}$ generated using the aforementioned framework.
A summary is provided in Table~\ref{tab:hi_fi_database_parameters}.
The total number of points, $N_{\mathrm{DB}} \in \mathbb{N}$, ranges from $524$ to $967$ per position.

\begin{table*}[tb]
\centering
\begin{tabular}{c|c|c|c|c|c|c|c|c|c}
\hline
\hline
$\hat{\rho}$ & $N_{\mathrm{DB}}$ & $\uptau_{\mathrm{min}}$ & $\uptau_{\mathrm{max}}$ & $\eta_{e, \mathrm{min}}$ & $\eta_{e, \mathrm{max}}$ & $\omega_{T_e, \mathrm{min}}$ & $\omega_{T_e, \mathrm{max}}$ & $Q_{e, \mathrm{min}}/Q_{\mathrm{GB}}$ & $Q_{e, \mathrm{max}}/Q_{\mathrm{GB}}$ \\
\hline
$0.2$ & $729$ & $0.93$ & $3.72$ & $1.25$ & $42.14$ & $0.75$ & $3.06$ & $0.02$ & $30.77$\\
$0.3$ & $822$ & $0.87$ & $3.40$ & $1.34$ & $32.48$ & $0.47$ & $3.50$ &  $0.02$ & $49.06$\\
$0.4$ & $674$ & $0.85$ & $2.44$ & $1.62$ & $40.55$ & $0.85$ & $3.75$ & $0.19$ & $80.41$\\
$0.5$ & $848$ & $0.84$ & $1.94$ & $1.69$ & $29.51$ & $1.53$ & $3.95$ &  $0.46$ & $94.84$\\
$0.6$ & $715$ & $0.90$ & $1.62$ & $1.04$ & $17.12$ & $1.55$ & $5.00$ &  $0.02$ & $116.71$\\
$0.7$ & $524$ & $0.99$ & $1.37$ & $2.03$ & $18.47$ & $1.94$ & $5.15$ &  $2.01$ & $136.93$\\
$0.8$ & $967$ & $1.01$ & $1.52$ & $1.46$ & $18.00$ & $1.15$ & $13.92$ &  $0.70$ & $415.06$\\
\hline
\hline
\end{tabular}
\caption{Database for $\hat{\rho} \in \{0.2, 0.3, 0.4, 0.5, 0.6, 0.7, 0.8 \}$ used for iterative model refinement and numerical validation.}
\label{tab:hi_fi_database_parameters}
\end{table*}

Since the primary objective of~\citet{Fernando2025} was code validation and profile prediction, the simulation database was generated as a byproduct rather than being specifically designed for reduced model development. 
The exploration of parameter space was governed by the \texttt{\textsc{Gene}-KNOSOS-Tango} framework, meaning that the simulations were not generated according to a dedicated sampling strategy and that the iterative procedure could not be steered toward prescribed parameter combinations.
It is also important to note that the resulting database is not guaranteed to provide a fully representative sampling of the W7-X operational parameter space. 
The explored parameter ranges depend strongly on the initial profile guesses used to initialize the coupled transport iterations. 
When the initial profiles already approximately satisfy the power and particle balances, \texttt{Tango} typically produces updated profiles that remain relatively close to the initial state, leading to a comparatively localized exploration of parameter space. 
In contrast, less accurate initial guesses can drive the profile evolution through a substantially broader region of parameter space, potentially extending beyond conditions realized in present experimental discharges. 
Because the database considered here was generated primarily using the former strategy, some parameters span a narrower range than is observed experimentally.

The sparse-grid-generated data were intended to complement the database at each $\hat{\rho}$ by providing additional samples within the experimentally relevant parameter space. 
In practice, this requires the sparse-grid boundaries to be consistent with experimentally accessible parameter ranges. 
However, at the time this study was initiated, sufficiently detailed parameter bounds were not available, and fixed bounds were therefore adopted based on the best available estimates. 
As a result, the sparse-grid extension includes regions of parameter space that are not represented in the simulation database and that, in some cases, may not correspond to experimentally explored operating conditions. 
This is particularly evident at $\hat{\rho} = 0.7$, where the range of $\uptau$ represented in the available database is substantially narrower than that observed experimentally~\cite{Wappl2025}, while the sparse-grid sampling extends beyond the experimentally relevant regime. 
Consequently, the bootstrap-based quantities used during the active learning procedure are partially influenced by parameter regions introduced through the sparse-grid initialization that extend beyond those typically encountered in W7-X operation. 
A more detailed discussion is provided in Appendix~\ref{sec_app:active_learning_sensitivity}.

\subsection{Database for three additional radial locations} \label{subsec:database_three_additional_loc}
For a more comprehensive perspective, we also construct radial-position–dependent coefficients for the scaling law~\eqref{eq:target_model} by performing regression on the coefficients of the reduced models for the original seven radial positions. 
The predictive performance of the resulting models is then assessed at three additional positions: $\hat{\rho} \in \{0.1, 0.55, 0.75\}$.
The data for these three locations were also generated by the \texttt{\textsc{Gene}-KNOSOS-Tango} framework.~\cite{Fernando2025} No \textsc{Gene} simulation data are available for $\hat{\rho} > 0.8$ in the current database, so model predictions beyond this point cannot be validated and extrapolation outside the simulated radial domain is not attempted here.
The available data are summarized in Table~\ref{tab:hi_fi_database_other_rad_loc}.

\begin{table*}[tb]
\centering
\begin{tabular}{c|c|c|c|c|c|c|c|c|c}
\hline
\hline
$\hat{\rho}$ & $N_{\mathrm{DB}}$ & $\uptau_{\mathrm{min}}$ & $\uptau_{\mathrm{max}}$ & $\eta_{e, \mathrm{min}}$ & $\eta_{e, \mathrm{max}}$ & $\omega_{T_e, \mathrm{min}}$ & $\omega_{T_e, \mathrm{max}}$ & $Q_{e, \mathrm{min}}/Q_{\mathrm{GB}}$ & $Q_{e, \mathrm{max}}/Q_{\mathrm{GB}}$ \\
\hline
$0.1$ & $251$ & $0.98$ & $3.85$ & $1.53$ & $17.75$ & $0.58$ & $2.46$ & $0.02$ & $15.97$\\
$0.55$ & $70$ & $1.10$ & $1.29$ & $4.79$ & $11.44$ & $1.95$ & $2.63$ &  $13.82$ & $41.42$\\
$0.75$ & $131$ & $1.02$ & $1.19$ & $2.41$ & $4.06$ & $2.54$ & $5.34$ & $9.52$ & $47.40$ \\
\hline
\hline
\end{tabular}
\caption{Database for three addition radial locations $\hat{\rho} \in \{0.1, 0.55, 0.75\}$ used to test generalized models.}
\label{tab:hi_fi_database_other_rad_loc}
\end{table*}

\section{Active-learning-based iterative reduced model refinement} \label{sec:ML_based_reduced_modeling}
To determine the coefficients $\{c_0, p_1, p_2, p_3\}$ of the target reduced models~\eqref{eq:target_model} for each of the seven $\hat{\rho}$, we employ an ML-driven approach that combines regression in logarithmic coordinates, similar to Ref.~\citenum{FMJ24}, with iterative refinement based on active learning. 
For each $\hat{\rho}$, we iteratively construct the training dataset, $\mathcal{P}_{\mathrm{train}}$, consisting of input-output pairs  $(\{\omega_{T_e}, \eta_e, \uptau\}, Q_{e}/Q_{\mathrm{GB}})$, where the explicit dependency on $\hat{\rho}$ is omitted to simplify the notation. 
The cardinality of this dataset is denoted by $N_{\mathrm{train}} \in \mathbb{N}$. 
At each iteration, $\mathcal{P}_{\mathrm{train}}$ is progressively enriched by adding the most informative input triplet 
$\{\omega_{T_e}, \eta_e, \uptau\}$ along with the corresponding heat flux value $Q_{e}/Q_{\mathrm{GB}}$, referred to as the new `label' in an ML context. 
The detailed steps of this procedure are described below.

\paragraph*{Step I: Training the initial reduced models.}
We begin by constructing initial (base) reduced models~\eqref{eq:target_model}. 
For each $\hat{\rho}$, the initial training dataset $\mathcal{P}_{\mathrm{train}}$ solely consists of the $N_{\mathrm{SG}}$ sparse grid data pairs $\{ (\{\omega_{T_e}, \eta_e, \uptau \}_i, Q_{e, i}/Q_{\mathrm{GB}}) \}_{i=1}^{N_{\mathrm{SG}}}$ from Sec.~\ref{subsec:SG_initmodel}. 
We then apply regression in logarithmic coordinates using this initial $\mathcal{P}_{\mathrm{train}}$ to determine the coefficients of the base models for each $\hat{\rho}$.
These models, presented in Eq.~\ref{eq:red_models_init}, show a strong dependence on $\omega_{T_e}$, indicated by the large values of $p_1$ across all radial locations.
In contrast, the dependence on $(\eta_e - 1)$ and $\uptau$ is less significant.
\begin{subequations}\label{eq:red_models_init}
    \begin{align}
        & \mathrm{\hat{\rho} = 0.2:}\quad Q_e/Q_{\mathrm{GB}} = 0.35 \, \omega_{T_e}^{3.08} \left(\eta_e - 1\right)^{0.41} \uptau^{-1.08} \label{eq:red_model_x0_0_2}                                                                            \\
        & \mathrm{\hat{\rho} = 0.3:}\quad Q_e/Q_{\mathrm{GB}} = 0.35 \, \omega_{T_e}^{2.90} \left(\eta_e - 1\right)^{0.51} \uptau^{-1.08} \label{eq:red_model_x0_0_3}                                                                            \\
        & \mathrm{\hat{\rho} = 0.4:}\quad Q_e/Q_{\mathrm{GB}} = 0.32 \, \omega_{T_e}^{2.99} \left(\eta_e - 1\right)^{0.48} \uptau^{-0.75} \label{eq:red_model_x0_0_4}                                                                            \\
        & \mathrm{\hat{\rho} = 0.5:}\quad Q_e/Q_{\mathrm{GB}} = 0.45 \, \omega_{T_e}^{2.80} \left(\eta_e - 1\right)^{0.48} \uptau^{-0.75} \label{eq:red_model_x0_0_5}                                                                            \\
        & \mathrm{\hat{\rho} = 0.6:}\quad Q_e/Q_{\mathrm{GB}} = 0.10 \, \omega_{T_e}^{3.39} \left(\eta_e - 1\right)^{0.78} \uptau^{-0.21} \label{eq:red_model_x0_0_6}                                                                            \\
        & \mathrm{\hat{\rho} = 0.7:}\quad Q_e/Q_{\mathrm{GB}} = 0.12 \, \omega_{T_e}^{3.28} \left(\eta_e - 1\right)^{0.90} \uptau^{-0.25} \label{eq:red_model_x0_0_7}                                                                            \\
        & \mathrm{\hat{\rho} = 0.8:}\quad Q_e/Q_{\mathrm{GB}} = 0.13 \, \omega_{T_e}^{3.31} \left(\eta_e - 1\right)^{0.79} \uptau^{-0.13} \label{eq:red_model_x0_0_8}
    \end{align}
\end{subequations}

\paragraph*{Step II: Measuring the models' prediction uncertainty on an input testing set.}
At each active learning step, we estimate the prediction uncertainty of the reduced models using bootstrapping~\cite{EB86}. 
To this end, we construct test sets of representative input parameter values, $\mathcal{I}_{\mathrm{test}}$, for each radial location; an adequate coverage of the input space is crucial for the success of the active learning procedure. 

\begin{figure*}
    \centering
    \includegraphics[width=0.99\textwidth]{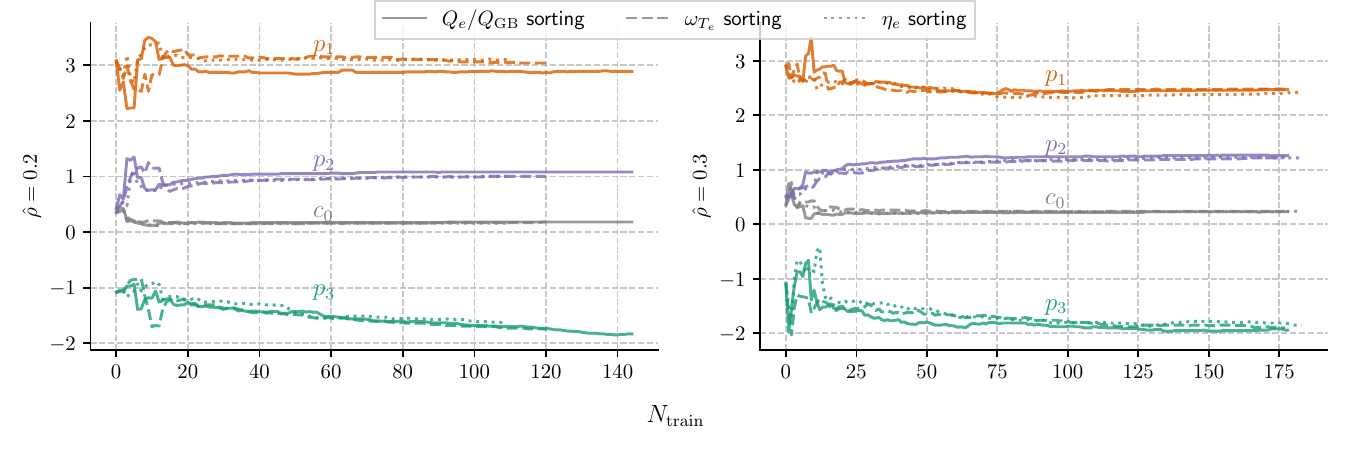}
    \caption{\label{fig:coeff_ev_x0_0_2_and_0_3}
        Evolution of the fitted coefficients $\{c_0(\hat{\rho}), p_1(\hat{\rho}), p_2(\hat{\rho}), p_3(\hat{\rho})\}$ for $\hat{\rho} \in \{0.2, 0.3\}$ during the active learning procedure for three strategies for choosing the testing set: sorting the entries of the available database with respect to $Q_e/Q_{\mathrm{GB}}, \omega_{T_e}$ or $\eta_e$. The procedure stops once the thresholds $\mathrm{max\_dev}$ and $\mathrm{max\_rel\_err}$ fall below $0.10$. }
\end{figure*}

\begin{figure*}
    \centering
    \includegraphics[width=0.99\textwidth]{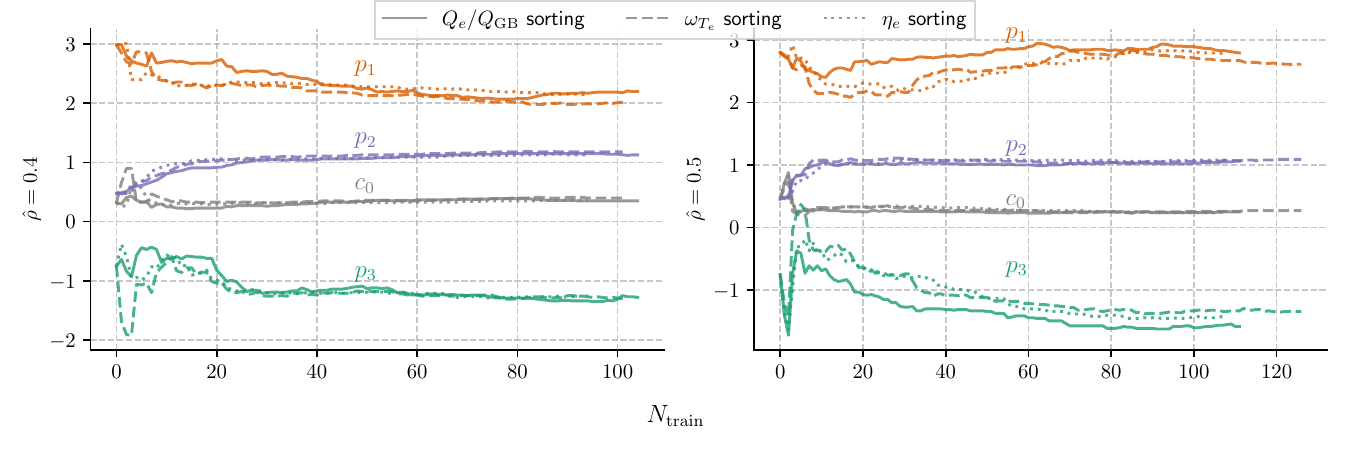}
    \caption{\label{fig:coeff_ev_x0_0_4_and_0_5}
        Evolution of the fitted coefficients $\{c_0(\hat{\rho}), p_1(\hat{\rho}), p_2(\hat{\rho}), p_3(\hat{\rho})\}$ for $\hat{\rho} \in \{0.4, 0.5\}$ during the active learning procedure for three strategies for choosing the testing set: sorting the entries of the available database with respect to $Q_e/Q_{\mathrm{GB}}, \omega_{T_e}$ or $\eta_e$. The procedure stops once the thresholds $\mathrm{max\_dev}$ and $\mathrm{max\_rel\_err}$ fall below $0.10$. }
\end{figure*}

\begin{figure*}
    \centering
    \includegraphics[width=0.99\textwidth]{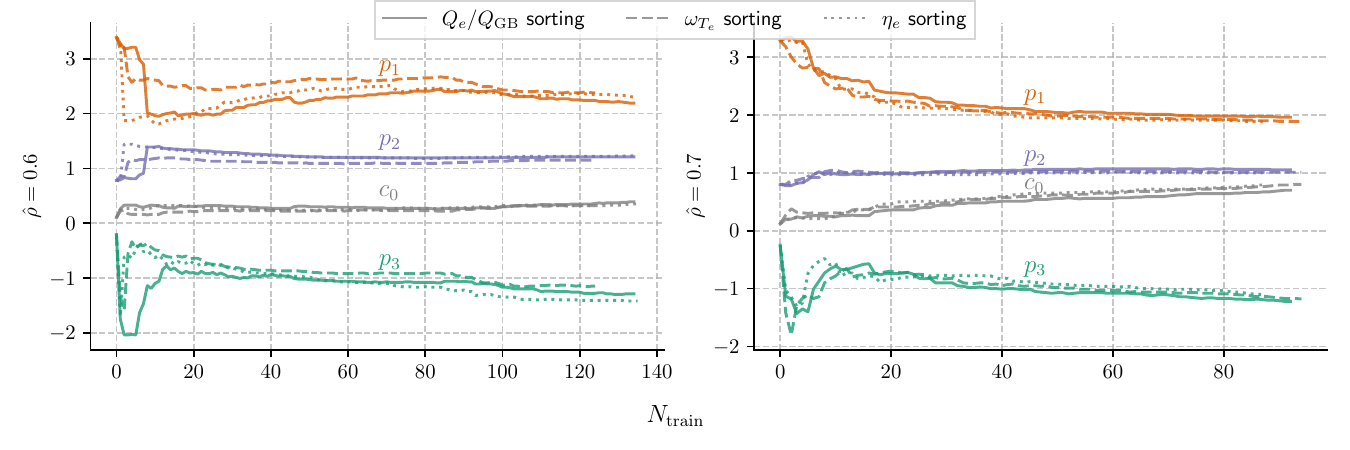}
    \caption{\label{fig:coeff_ev_x0_0_6_and_0_7}
        Evolution of the fitted coefficients $\{c_0(\hat{\rho}), p_1(\hat{\rho}), p_2(\hat{\rho}), p_3(\hat{\rho})\}$ for $\hat{\rho} \in \{0.6, 0.7\}$ during the active learning procedure for three strategies for choosing the testing set: sorting the entries of the available database with respect to $Q_e/Q_{\mathrm{GB}}, \omega_{T_e}$ or $\eta_e$. The procedure stops once the thresholds $\mathrm{max\_dev}$ and $\mathrm{max\_rel\_err}$ fall below $0.10$. }
\end{figure*}

\begin{figure}
    \centering
    \includegraphics[width=0.49\textwidth]{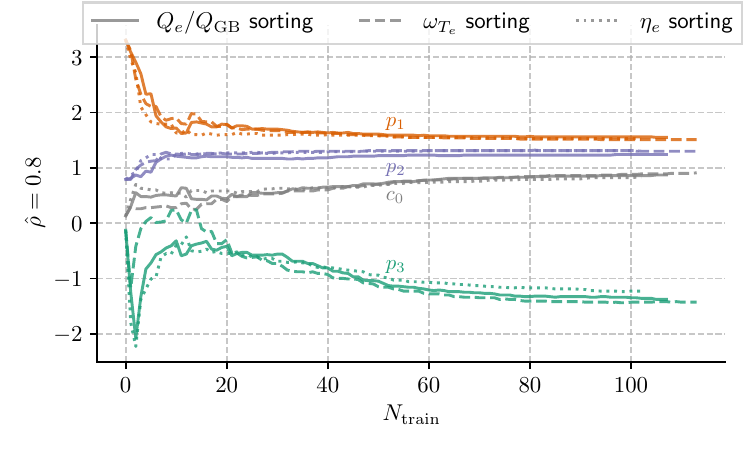}
    \caption{\label{fig:coeff_ev_x0_0_8}
        Evolution of the fitted coefficients $\{c_0(\hat{\rho}), p_1(\hat{\rho}), p_2(\hat{\rho}), p_3(\hat{\rho})\}$ for $\hat{\rho} = 0.8$ during the active learning procedure for three strategies for choosing the testing set: sorting the entries of the available database with respect to $Q_e/Q_{\mathrm{GB}}, \omega_{T_e}$ or $\eta_e$. The procedure stops once the thresholds $\mathrm{max\_dev}$ and $\mathrm{max\_rel\_err}$ fall below $0.10$. }
\end{figure}

These test sets are derived from the pre-existing database summarized in Sec.~\ref{subsec:hi_fi_database}.
Specifically, all database entries are sorted in ascending order of heat flux.
The set $\mathcal{I}_{\mathrm{test}}$ is then constructed by selecting every fourth input triplet $\{\omega_{T_e}, \eta_e, \uptau\}$ from the sorted database.
Consequently, each test set contains approximately one quarter of the database entries, i.e., on the order of $N_{\mathrm{DB}}/4$ samples per $\hat{\rho}$.
While this construction does not enforce proximity in the input space between successive samples, we observe that the resulting active learning behavior is qualitatively robust to the ordering choice, as illustrated in Figs.~\ref{fig:coeff_ev_x0_0_2_and_0_3}--\ref{fig:coeff_ev_x0_0_8}.
Furthermore, additional tests based on pseudo-random selections of the test set show qualitatively consistent results, suggesting limited sensitivity to the specific choice of $\mathcal{I}_{\mathrm{test}}$.
The remaining database pairs $(\{\omega_{T_e}, \eta_e, \uptau\}, Q_{e}/Q_{\mathrm{GB}})$, denoted by $\mathcal{P}_{\mathrm{pred}}$, are reserved for independent numerical validation; the corresponding sample counts $N_{\mathrm{pred}} \in \mathbb{N}$ are listed in the last column of Table~\ref{tab:final_values_ntrain_nval}.

Next, we employ bootstrapping to construct an ensemble of $M$ models for each $\hat{\rho}$. We resample the current training dataset, $\mathcal{P}_{\mathrm{train}}$, with replacement $M$ times to generate $M$ bootstrap datasets of identical size.
Each model is refitted on its corresponding dataset and used to compute predictions for all input values in $\mathcal{I}_{\mathrm{test}}$. 
This procedure yields an ensemble of $M$ predictions for every input triplet in the test set, providing an empirical estimate of the prediction distribution and the uncertainty induced by variability in the training data.

We then quantify prediction uncertainty using these ensembles. 
For each test point, we compute the $2.5$th and $97.5$th percentiles of the bootstrap predictions, which define the lower and upper bounds of the $95\%$ prediction interval. We also compute the ensemble mean prediction.
As the size of the training dataset increases, these estimates typically stabilize and the prediction intervals tend to narrow. 
We note, however, that narrow bootstrap intervals do not necessarily imply physically accurate predictions; rather, they reflect confidence in the fitted model, within the assumed functional form and least-squares framework.
Although the interval width quantifies variability across the bootstrap ensemble, we instead measure uncertainty via the relative deviation of the interval bounds from the mean prediction. 
This normalization renders the uncertainty metric independent of the wide dynamic range of the heat flux values (from $\sim 0.2$ to over $400$; see Table~\ref{tab:hi_fi_database_parameters}).
To ensure robust estimation, we use $M = 10{,}000$ bootstrap samples in all experiments.

\paragraph*{Step III: Active-learning-based model refinement.}
The active learning procedure selects the input triplet $\{\omega_{T_e}, \eta_e, \uptau \}$ from the test set that has the largest deviation of the prediction interval bounds from the mean prediction.
Once selected, this input is removed from $\mathcal{I}_{\mathrm{test}}$ and is added to the training dataset, $\mathcal{P}_{\mathrm{train}}$, along with its corresponding heat flux value (or ``label" in ML terminology). 
The models are then retrained on this augmented dataset, completing one cycle.
We note that although active learning is typically performed online, where $\mathcal{I}_{\mathrm{test}}$ might consist of a large number of pseudo-random samples of inputs for example, our use of the pre-existing database is a pragmatic choice made to demonstrate the method's efficacy within the constraints of the available datasets.
The goal of our procedure is to efficiently and robustly identify the `best' models that the available database can support, relative to the guiding active learning metrics.

\paragraph*{Step IV: Convergence and final model construction.}
We repeat Steps II and III until the maximum relative deviation of the prediction interval bounds from the mean falls below a user-defined threshold. 
However, this criterion alone may be insufficient, as a small deviation does not guarantee that the predicted values are close to the true heat flux. 
To address this, we introduce a second criterion: the relative error between each newly added heat flux value and its prediction from the model at the previous active learning step must also remain below a user-defined threshold. 
The procedure terminates only when both the maximum deviation and the relative error satisfy their respective thresholds, denoted by $\mathrm{max\_dev} > 0$ and $\mathrm{max\_rel\_err} > 0$. 
These thresholds can be tightened during the procedure if higher accuracy is required. 
Upon termination, the resulting training dataset of size $N_{\mathrm{train}}$ is used to construct the final model for each $\hat{\rho}$ via regression.

Figures~\ref{fig:coeff_ev_x0_0_2_and_0_3}--\ref{fig:coeff_ev_x0_0_8} illustrate the convergence of the fitted coefficients for $\mathrm{max\_dev} = \mathrm{max\_rel\_err} = 0.10$, demonstrating that the active learning procedure is not overly sensitive to the choice of testing-set sorting strategy (i.e., sorting by $Q_e/Q_{\mathrm{GB}}$, $\omega_{T_e}$, or $\eta_e$). 
Moreover, the fitted coefficients remain relatively stable as the tolerances approach the final value of $0.10$, indicating that the active-learning procedure has converged by this stage. 
Since the testing set is fixed while the training sets are cumulative as the tolerances decrease, stricter values of \texttt{max\_dev} and \texttt{max\_rel\_err} naturally produce progressively larger training sets, with the models obtained at $0.10$ already incorporating all samples selected at larger tolerance values. 
Additional experiments further showed that decreasing the tolerances below $0.10$ yields only marginal improvements in predictive accuracy while increasing the size of the training sets. 
Accordingly, the final models adopt $\mathrm{max\_dev} = \mathrm{max\_rel\_err} = 0.10$.

\begin{table}
\centering
\begin{tabular}{c|c|c}
\hline
\hline
$\hat{\rho}$ & $N_{\mathrm{train}}$ & $N_{\mathrm{pred}}$\\
\hline
$0.2$ & $155$ & $546$\\
$0.3$ & $190$ & $616$\\
$0.4$ & $115$ & $505$\\
$0.5$ & $122$ & $636$\\
$0.6$ & $146$ & $536$\\
$0.7$ & $104$ & $393$\\
$0.8$ & $119$ & $725$\\
\hline
\hline
\end{tabular}
\caption{The cardinality of the training dataset ($\mathrm{N_{train}}$) at the end of the iterative model refinement procedure and the number of validation data pairs ($\mathrm{N_{\mathrm{pred}}}$).}
\label{tab:final_values_ntrain_nval}
\end{table}

The fitted reduced models are presented in Eq.~\eqref{eq:red_models_final}. 
The signs of the exponents $\{p_1, p_2, p_3\}$ indicate the influence of the three model parameters on the ETG heat flux. 
The normalized electron temperature radial gradient, $\omega_{T_e}$, is a primary driver of the instability, which explains the positive exponent of this term. 
The parameter $\eta_e$ (i.e., the ratio of $\omega_{T_e}$ to $\omega_{n_e}$) captures competing effects: the temperature gradient destabilizes ETG modes, while the density gradient has a stabilizing effect. 
Consequently, a positive exponent is expected. 
Increasing the temperature ratio, $\uptau$, has a stabilizing effect on ETG modes, which results in a negative exponent for this variable. 
Examining the radial trends, the exponent of $\eta_e - 1$ varies little with $\hat{\rho}$, indicating a nearly uniform radial dependency. 
In contrast, the exponent for $\uptau$ (i.e., $p_3$) generally decreases in absolute value toward the W7-X edge, consistent with the observation that $\uptau$ exhibits larger variations near the core and smaller variations near the edge. 
Similarly, the exponent of $\omega_{T_e}$ tends to decrease with increasing $\hat{\rho}$.
We note that because we employ $M = 10{,}000$ bootstrap samples, the estimation of coefficient variability is well converged, and the resulting fitted models are robust with respect to changes in the random seed in the active learning procedure.
\begin{subequations}\label{eq:red_models_final}
    \begin{align}
        & \mathrm{\hat{\rho} = 0.2:}\quad Q_e/Q_{\mathrm{GB}} = 0.18 \, \omega_{T_e}^{2.89} \left(\eta_e - 1\right)^{1.08} \uptau^{-1.83} \label{eq:red_model_x0_0_2}                                                                            \\
        & \mathrm{\hat{\rho} = 0.3:}\quad Q_e/Q_{\mathrm{GB}} = 0.23 \, \omega_{T_e}^{2.47} \left(\eta_e - 1\right)^{1.26} \uptau^{-1.94} \label{eq:red_model_x0_0_3}                                                                            \\
        & \mathrm{\hat{\rho} = 0.4:}\quad Q_e/Q_{\mathrm{GB}} = 0.35 \, \omega_{T_e}^{2.20} \left(\eta_e - 1\right)^{1.13} \uptau^{-1.28} \label{eq:red_model_x0_0_4}                                                                            \\
        & \mathrm{\hat{\rho} = 0.5:}\quad Q_e/Q_{\mathrm{GB}} = 0.25 \, \omega_{T_e}^{2.79} \left(\eta_e - 1\right)^{1.06} \uptau^{-1.58} \label{eq:red_model_x0_0_5}                                                                            \\
        & \mathrm{\hat{\rho} = 0.6:}\quad Q_e/Q_{\mathrm{GB}} = 0.39 \, \omega_{T_e}^{2.19} \left(\eta_e - 1\right)^{1.21} \uptau^{-1.28} \label{eq:red_model_x0_0_6}                                                                            \\
        & \mathrm{\hat{\rho} = 0.7:}\quad Q_e/Q_{\mathrm{GB}} = 0.70 \, \omega_{T_e}^{1.96} \left(\eta_e - 1\right)^{1.05} \uptau^{-1.22} \label{eq:red_model_x0_0_7}                                                                            \\
        & \mathrm{\hat{\rho} = 0.8:}\quad Q_e/Q_{\mathrm{GB}} = 0.87 \, \omega_{T_e}^{1.55} \left(\eta_e - 1\right)^{1.24} \uptau^{-1.38} \label{eq:red_model_x0_0_8}
    \end{align}
\end{subequations}

Comparing our reduced models with corresponding terms in ETG reduced models for tokamak pedestals \cite{FMJ24,Ha22,Hatch_2024}, we note the following. 
The analogous term in \citet{FMJ24} is $\omega_{T_e}^{1.40} (\eta_e - 1)^{1.79} \uptau^{-0.76}$, while the model from \citet{Ha22} contains $\omega_{T_e}\left(1.44+0.50 \eta_e^4\right)$, and the more recent model in \citet{Hatch_2024} uses $\omega_{T_e}^{2} (\eta_e - 1)\, \eta_e^{1.54} \uptau^{-0.5}$. 
Since these studies fitted models only for the tokamak pedestal, radial trends are unavailable, and direct comparison across $\hat{\rho}$ is not possible. 
Nevertheless, we can compare their characteristics with Eq.~\ref{eq:red_model_x0_0_8}, i.e., our outermost model at $\hat{\rho} = 0.8$. 
The exponent of $\omega_{T_e}$ in Eq.~\ref{eq:red_model_x0_0_8} is comparable in magnitude to those in the aforementioned ETG pedestal models. 
Similarly, the exponents of $(\eta_e - 1)$, when present, are of similar order. 
Beyond differences in functional form and parameter inclusion, the primary distinction is that Eq.~\ref{eq:red_models_final} predicts a stronger stabilizing effect due to $\uptau$. 
Comparing these ETG pedestal models with our innermost reduced model, Eq.~\ref{eq:red_model_x0_0_2}, further highlights discrepancies in the magnitudes of the exponents. 
These observations underscore both similarities and differences between our stellarator models and tokamak pedestal models.
Recent reduced models for core ETG turbulence in tokamaks, however, adopt a different functional form. 
For example, the reduced model for ETG heat diffusivity in \citet{Muraca2023}, which can be related to heat flux, features a threshold in $\omega_{T_e}$ rather than $\eta_e$. 
This reflects the predominantly toroidal character of ETG turbulence in the tokamak core. 
In contrast, ETG turbulence in our W7-X simulation database retains a substantial slab-like component, consistent with the $\eta_e$-based threshold of Eq.~\ref{eq:red_models_final}.

\begin{figure}
    \centering
    \includegraphics[width=0.49\textwidth]{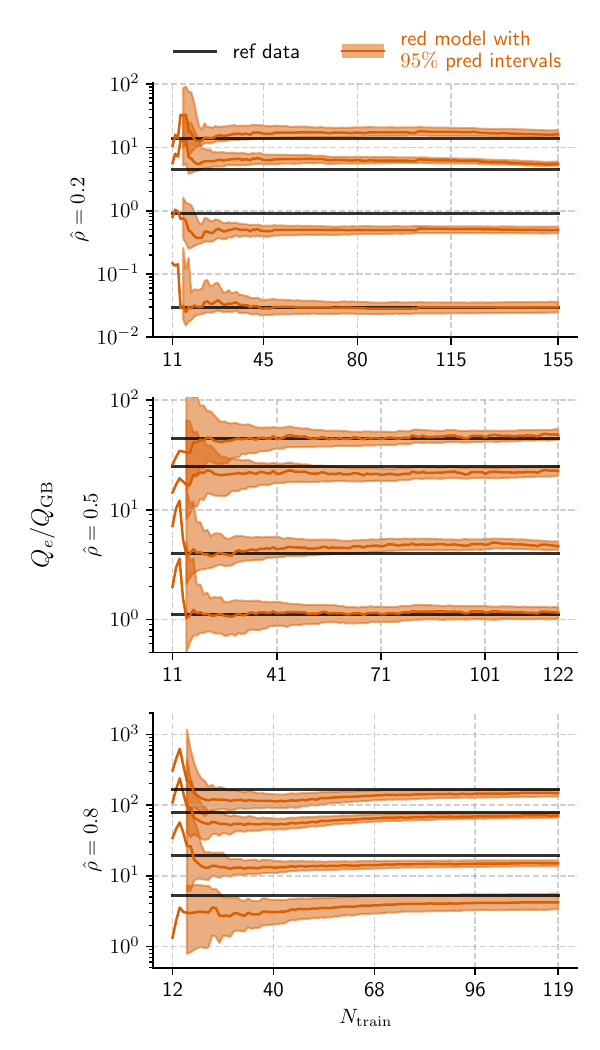}
    \caption{\label{fig:red_model_pred_x0_0_2_to_0_8}
        Evolution of reduced model predictions for four validation heat fluxes as a function of $N_{\mathrm{train}}$ in the active learning procedure for $\hat{\rho} \in \{0.2, 0.5, 0.8\}$. 
        Shaded regions indicate 95\% prediction intervals; the very large intervals from the first few iterations are omitted for clarity.
}
\end{figure}

Figure~\ref{fig:red_model_pred_x0_0_2_to_0_8} shows the evolution of reduced model predictions as the training set size increases during the active learning procedure. 
Results are shown for four validation heat flux values (from small to large) at $\hat{\rho} \in \{0.2, 0.5, 0.8\}$. 
Prediction uncertainty is quantified via $95\%$ bootstrapped intervals; the extremely large intervals from the first four iterations are omitted for clarity. 
As more informative training data are added, predictions stabilize and interval widths decrease. 
Recall that narrow intervals indicate confidence in the fitted model within the assumed functional form and least-squares framework, but do not necessarily guarantee physically accurate predictions.

\begin{table*}
\centering
\begin{tabular}{c|cccc|cccc|cccc|cccc}
\hline
\hline
$\hat{\rho}$ & $c_{0, \mathrm{CI \ lo}}$ & $c_{0, \mathrm{CI \ up}}$ & $\bar{c}_{0, \mathrm{CI}}$ & $c_{0, \mathrm{det}}$ & $p_{1, \mathrm{CI \ lo}}$ & $p_{1, \mathrm{CI \ up}}$ & $\bar{p}_{1, \mathrm{CI}}$ & $p_{1, \mathrm{det}}$ & $p_{2, \mathrm{CI \ lo}}$ & $p_{2, \mathrm{CI \ up}}$ & $\bar{p}_{2, \mathrm{CI}}$ & $p_{2, \mathrm{det}}$ & $p_{3, \mathrm{CI \ lo}}$ & $p_{3, \mathrm{CI \ up}}$ & $\bar{p}_{3, \mathrm{CI}}$ & $p_{3, \mathrm{det}}$\\
\hline
$0.2$ & $0.15$ & $0.21$ & $0.18$ & $0.18$ & $2.52$ & $3.26$ & $2.89$ & $2.89$ & $0.94$ & $1.25$ & $1.08$ & $1.08$ & $-2.09$ & $-1.63$ & $-1.83$ & $-1.83$\\
$0.3$ & $0.21$ & $0.25$ & $0.23$ & $0.23$ & $2.31$ & $2.61$ & $2.47$ & $2.47$ & $1.14$ & $1.42$ & $1.26$ & $1.26$ & $-2.23$ & $-1.68$ & $-1.94$ & $-1.94$\\
$0.4$ & $0.29$ & $0.43$ & $0.35$ & $0.35$ & $1.92$ & $2.44$ & $2.20$ & $2.20$ & $0.96$ & $1.34$ & $1.13$ & $1.13$ & $-1.84$ & $-0.81$ & $-1.28$ & $-1.28$\\
$0.5$ & $0.19$ & $0.31$ & $0.25$ & $0.25$ & $2.51$ & $3.15$ & $2.78$ & $2.79$ & $0.97$ & $3.15$ & $1.06$ & $1.06$ & $-2.26$ & $-1.17$ & $-1.58$ & $-1.58$\\
$0.6$ & $0.27$ & $0.59$ & $0.39$ & $0.39$ & $1.73$ & $2.61$ & $2.19$ & $2.19$ & $1.05$ & $1.37$ & $1.21$ & $1.21$ & $-2.01$ & $-0.56$ & $-1.28$ & $-1.29$\\
$0.7$ & $0.44$ & $1.43$ & $0.70$ & $0.70$ & $1.48$ & $2.29$ & $1.96$ & $1.96$ & $0.91$ & $1.17$ & $1.05$ & $1.05$ & $-1.86$ & $-0.59$ & $-1.23$ & $-1.22$\\
$0.8$ & $0.59$ & $1.34$ & $0.86$ & $0.87$ & $1.35$ & $1.73$ & $1.55$ & $1.55$ & $1.13$ & $1.35$ & $1.24$ & $1.24$ & $-2.20$ & $-0.87$ & $-1.37$ & $-1.38$\\
\hline
\hline
\end{tabular}
\caption{$95\%$ confidence intervals for the inferred model coefficients $\{c_0, p_1, p_2, p_3\}$. Intervals are computed using a percentile bootstrapping approach. For each coefficient, the table presents the lower ($2.5$th percentile) and upper ($97.5$th percentile) bound and the bootstrapped mean, as well as the point estimate (deterministic value) from the original training dataset.}
\label{tab:boostrap_reduced_model_cofficients}
\end{table*}

Finally, we assess the confidence intervals of the inferred model coefficients $\{c_0, p_1, p_2, p_3\}$ in Eq.~\ref{eq:red_models_final} using the percentile bootstrap approach with $M = 10{,}000$ samples.
This procedure retrains the models $M$ times by resampling the final active-learning training set with replacement.
The confidence intervals are then estimated from the $2.5\text{th}$ and $97.5\text{th}$ percentiles of the resulting parameter ensembles.
Table~\ref{tab:boostrap_reduced_model_cofficients} reports the resulting confidence intervals together with the corresponding deterministic coefficient estimates from Eq.~\ref{eq:red_models_final}.
The bootstrap means remain close to the deterministic estimates, indicating that the inferred coefficients are robust to resampling of the training dataset.
We emphasize, however, that these confidence intervals reflect only the sensitivity of the fitted coefficients to the specific training points selected during active learning, that is, how consistently the coefficients are estimated from the training dataset.
They do not assess whether the assumed functional form of the scaling law is correct, nor whether the model remains reliable outside the explored region of parameter space.
Tight confidence intervals should therefore not be interpreted as evidence of broad model validity, but rather as an indication that the fitted coefficients are statistically stable within the adopted modeling framework and the training dataset.


\section{Predictions using the proposed models}\label{sec:results_original_seven_loc}

We now assess the prediction capabilities of the proposed reduced models for $\hat{\rho} \in \{0.2, 0.3, 0.4, 0.5, 0.6, 0.7, 0.8 \}$ beyond training using the numerical validation datasets $\mathcal{P}_{\mathrm{pred}}$ described in Sec.~\ref{sec:ML_based_reduced_modeling}.
To evaluate the accuracy of the models' deterministic predictions, we use the error metric from~\citet{Ha22,Hatch_2024} and~\citet{FMJ24}: 
\begin{equation} \label{eq:error_measure}
\begin{split}
    \varepsilon\left(\frac{Q_{\mathrm{e, ref}}}{Q_{\mathrm{GB}}}, \frac{Q_{\mathrm{e, pred}}}{Q_{\mathrm{GB}}}\right) = \\ & \kern-5em \sqrt{\frac{1}{N_{\mathrm{pred}}} \sum_{i=1}^{N_{\mathrm{pred}}} \frac{\left(Q_{\mathrm{e, ref}; i}/Q_{\mathrm{GB}} - Q_{\mathrm{e, pred}; i}/Q_{\mathrm{GB}}\right)^2}{\left(Q_{\mathrm{e, ref}; i}/Q_{\mathrm{GB}} + Q_{\mathrm{e, pred}; i}/Q_{\mathrm{GB}}\right)^2}}\,,
\end{split}
\end{equation}
which compares the $N_{\mathrm{pred}}$ reference electron heat fluxes from the database, $Q_{\mathrm{e, ref}}/Q_{\mathrm{GB}}$, with their corresponding reduced model predictions, $Q_{\mathrm{e, pred}}/Q_{\mathrm{GB}}$.
This error equally penalizes extreme over- and under-prediction, that is, $Q_{\mathrm{e, pred}} \gg Q_{\mathrm{e, ref}}$ and $Q_{\mathrm{e, pred}} \ll Q_{\mathrm{e, ref}}$.
\begin{table}
\centering
\begin{tabular}{c|c|c}
\hline
\hline
$\hat{\rho}$ & $\varepsilon_{\mathrm{train}}$ & $\varepsilon_{\mathrm{pred}}$\\
\hline
$0.2$ & $0.172$ & $0.161$\\
$0.3$ & $0.180$ & $0.179$\\
$0.4$ & $0.153$ & $0.142$\\
$0.5$ & $0.150$ & $0.173$\\
$0.6$ & $0.195$ & $0.168$\\
$0.7$ & $0.135$ & $0.109$\\
$0.8$ & $0.167$ & $0.112$\\
\hline
\hline
\end{tabular}
\caption{Errors for training (second column) and numerical validation (third column) for all considered radial locations.}
\label{tab:training_and_validation_eps}
\end{table}

\begin{figure*}
    \centering
    \includegraphics[width=0.99\textwidth]{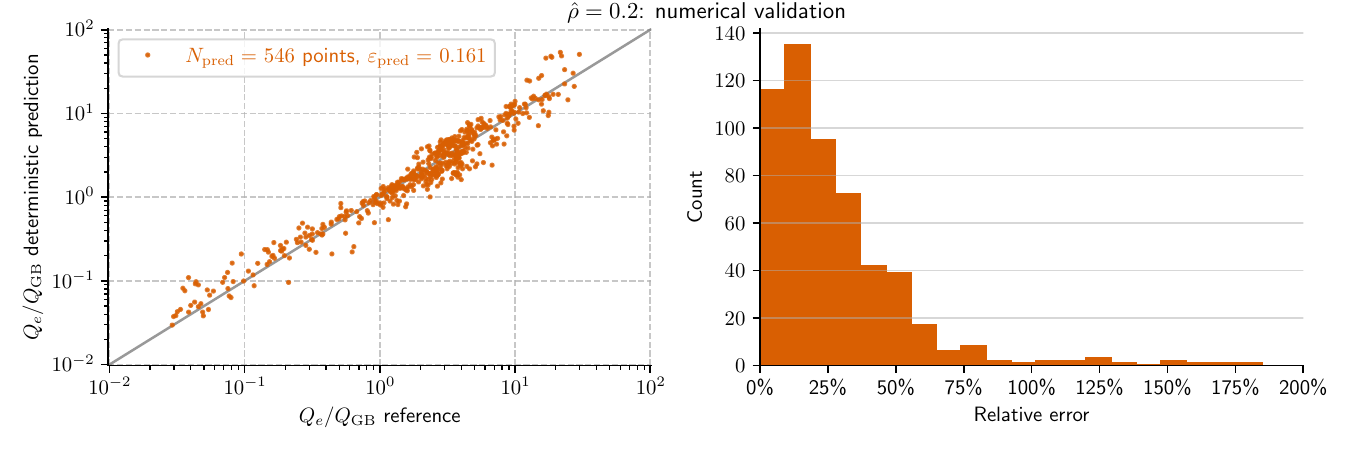}
    \caption{\label{fig:red_model_pred_x0_0_2}
        Numerical validation of the reduced model~\eqref{eq:red_model_x0_0_4} at $\hat{\rho} = 0.2$. The left plot compares its deterministic predictions with all $N_{\mathrm{pred}}=546$ reference values. The right plot shows the histogram of relative errors.}
\end{figure*}

\begin{figure*}
    \centering
    \includegraphics[width=0.99\textwidth]{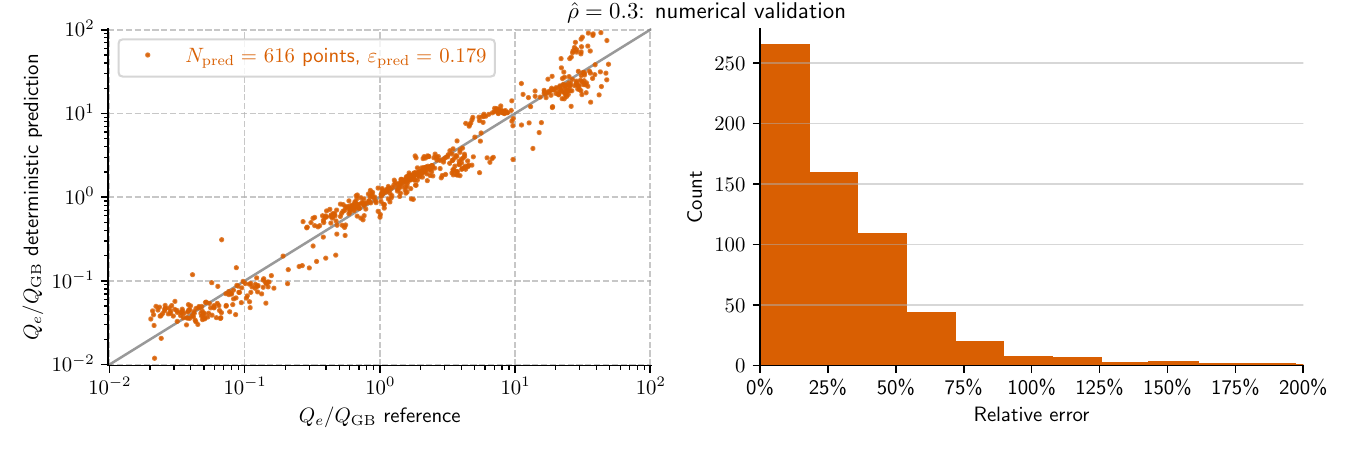}
    \caption{\label{fig:red_model_pred_x0_0_3}
        Numerical validation of the reduced model~\eqref{eq:red_model_x0_0_3} at $\hat{\rho} = 0.3$. The left plot compares its deterministic predictions with all $N_{\mathrm{pred}}=616$ reference values. The right plot shows the histogram of relative errors.}
\end{figure*}

\begin{figure*}
    \centering
    \includegraphics[width=0.99\textwidth]{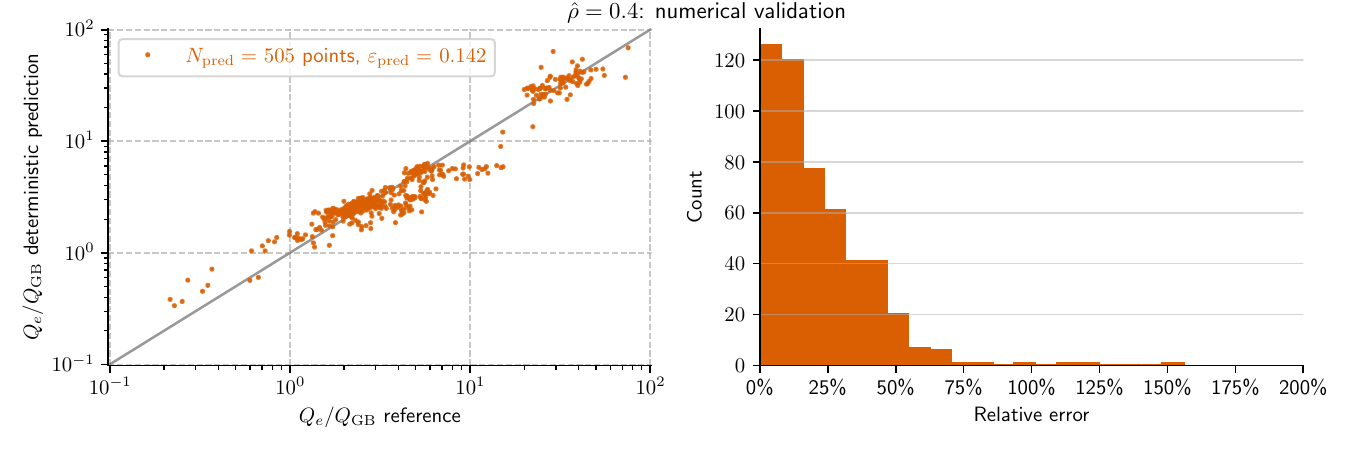}
    \caption{\label{fig:red_model_pred_x0_0_4}
        Numerical validation of the reduced model~\eqref{eq:red_model_x0_0_4} at $\hat{\rho} = 0.4$. The left plot compares its deterministic predictions with all $N_{\mathrm{pred}}=505$ reference values. The right plot shows the histogram of relative errors.}
\end{figure*}

\begin{figure*}
    \centering
    \includegraphics[width=0.99\textwidth]{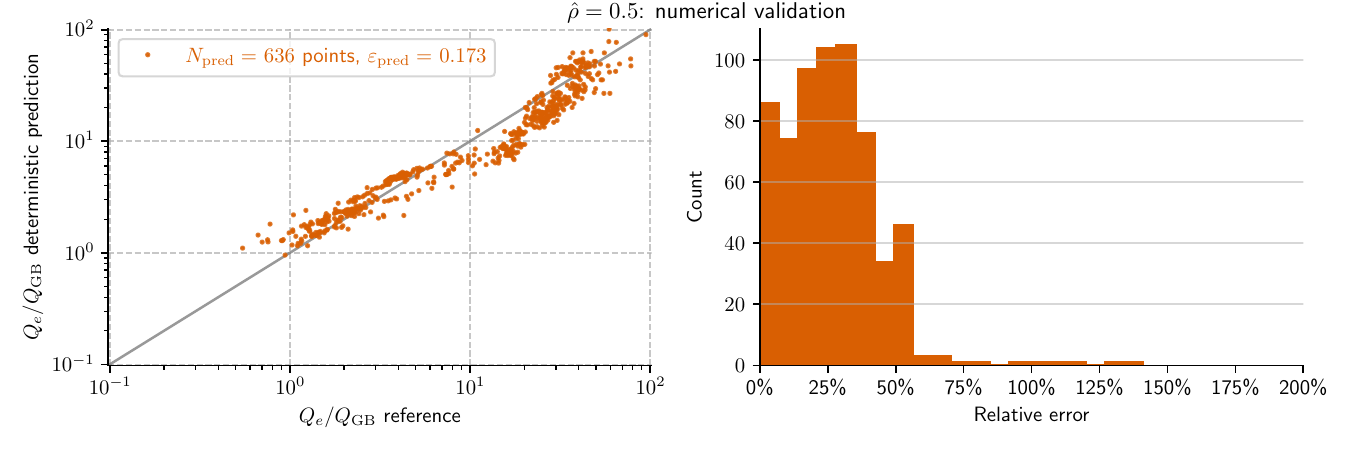}
    \caption{\label{fig:red_model_pred_x0_0_5}
        Numerical validation of the reduced model~\eqref{eq:red_model_x0_0_5} at $\hat{\rho} = 0.5$. The left plot compares its deterministic predictions with all $N_{\mathrm{pred}}=636$ reference values. The right plot shows the histogram of relative errors.}
\end{figure*}

\begin{figure*}
    \centering
    \includegraphics[width=0.99\textwidth]{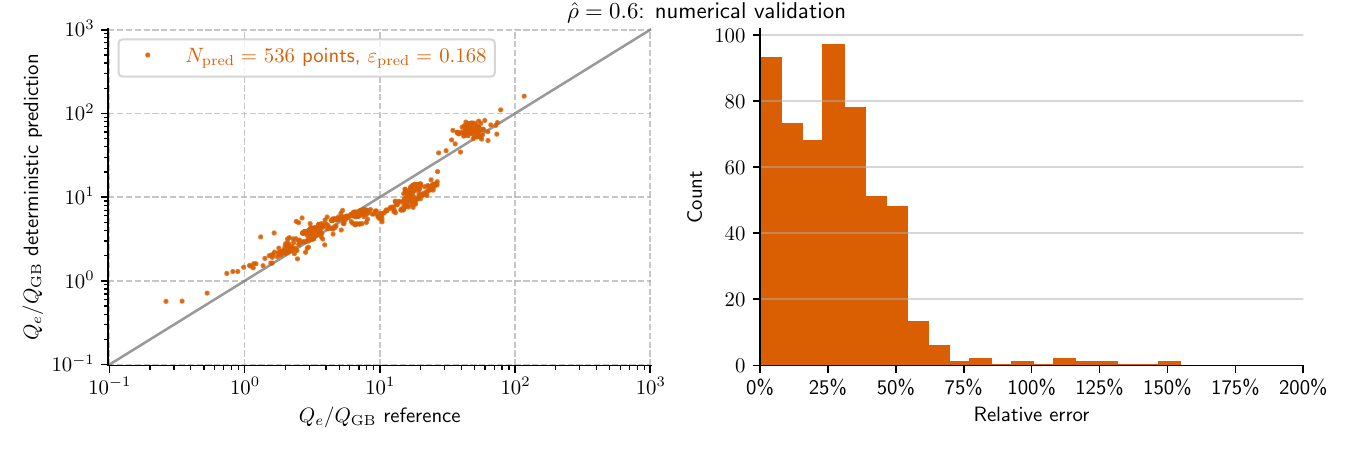}
    \caption{\label{fig:red_model_pred_x0_0_6}
        Numerical validation of the reduced model~\eqref{eq:red_model_x0_0_6} at $\hat{\rho} = 0.6$. The left plot compares its deterministic predictions with all $N_{\mathrm{pred}}=536$ reference values. The right plot shows the histogram of relative errors.}
\end{figure*}

\begin{figure*}
    \centering
    \includegraphics[width=0.99\textwidth]{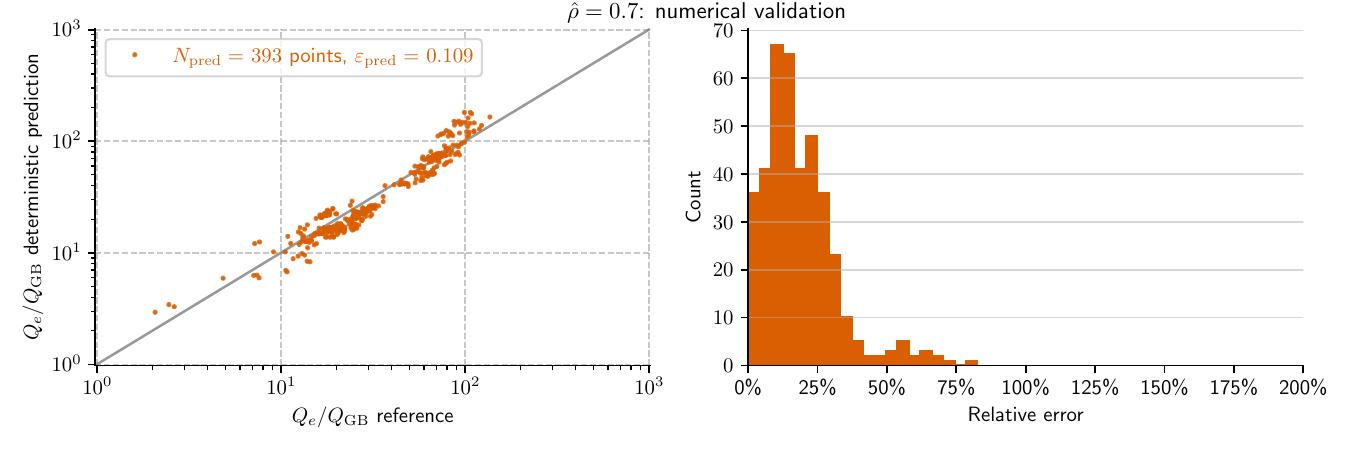}
    \caption{\label{fig:red_model_pred_x0_0_7}
        Numerical validation of the reduced model~\eqref{eq:red_model_x0_0_7} at $\hat{\rho} = 0.7$. The left plot compares its deterministic predictions with all $N_{\mathrm{pred}}=393$ reference values. The right plot shows the histogram of relative errors.}
\end{figure*}

\begin{figure*}
    \centering
    \includegraphics[width=0.99\textwidth]{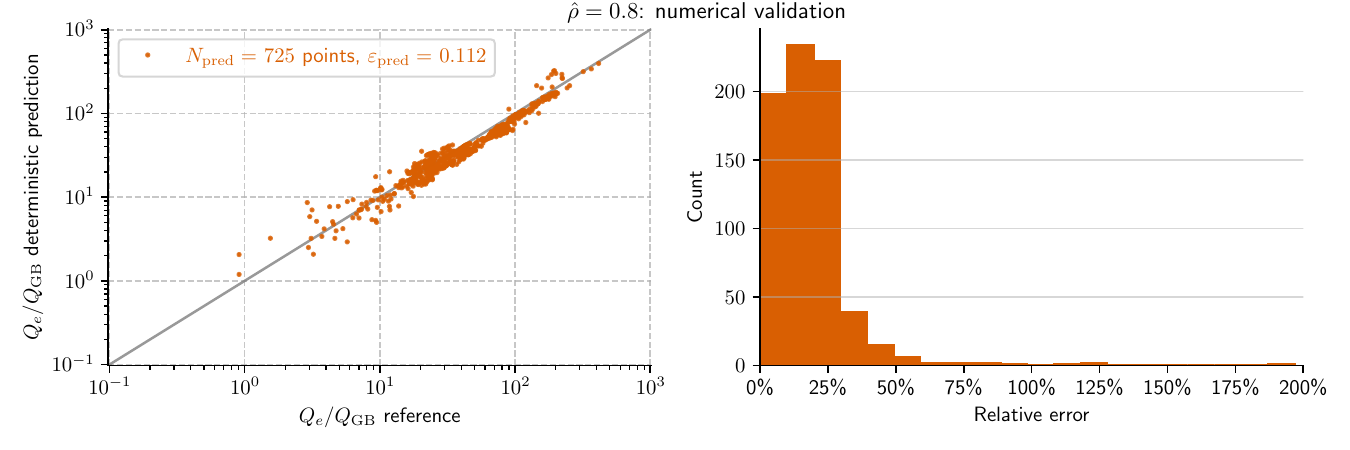}
    \caption{\label{fig:red_model_pred_x0_0_8}
        Numerical validation of the reduced model~\eqref{eq:red_model_x0_0_8} at $\hat{\rho} = 0.8$. The left plot compares its deterministic predictions with all $N_{\mathrm{pred}}=725$ reference values. The right plot shows the histogram of relative errors.}
\end{figure*}

\begin{figure*}
    \centering
    \includegraphics[width=0.99\textwidth]{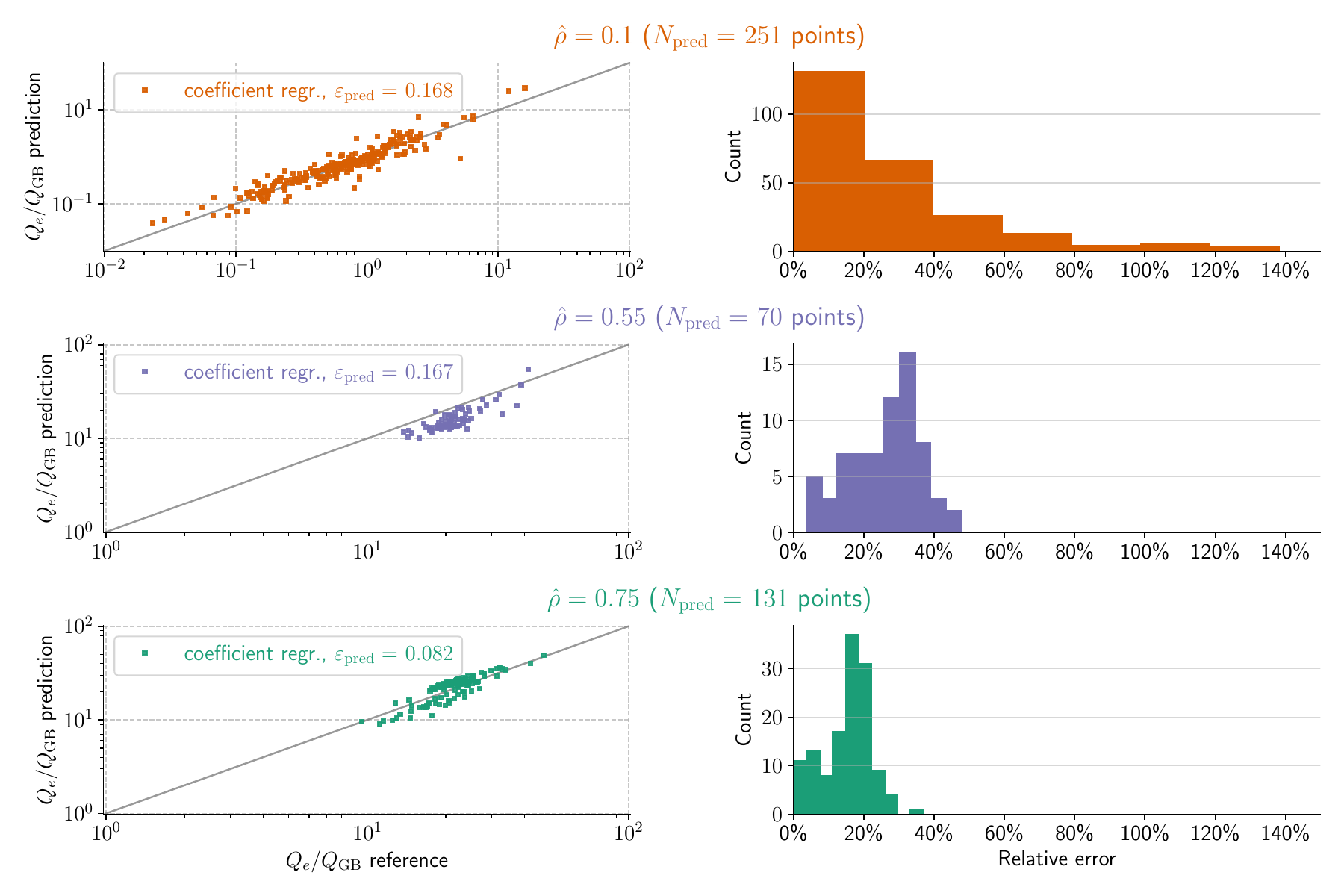}
    \caption{\label{fig:red_model_pred_x0_0_1_and_0_55_and_0_75}
        Left: Predictions for $\hat{\rho} = 0.1$ (top), $\hat{\rho} = 0.55$ (center), and $\hat{\rho} = 0.75$ (bottom) using models~\eqref{eq:red_models_x0_0_1_0_55_0_75} with coefficients obtained via the regression formulae~\eqref{eq:curve_fit_coefficients}. Right: histogram of relative errors.}
\end{figure*}

Table~\ref{tab:training_and_validation_eps} shows the errors for both training ($\varepsilon_{\mathrm{train}}$, second column) and numerical validation ($\varepsilon_{\mathrm{pred}}$, third column) for all seven $\hat{\rho}$. 
All $\varepsilon_{\mathrm{train}}$ values are below $0.20$, indicating that our iteratively refined reduced models~\eqref{eq:red_models_final} reproduce their training data with reasonable accuracy.
The true test, however, is their generalization capability, which we discuss next.

Figures~\ref{fig:red_model_pred_x0_0_2}--\ref{fig:red_model_pred_x0_0_8} plot the predictions from our reduced models across all seven radial locations. 
The left plots compare the $N_{\mathrm{pred}}$ predictions (from Eq.~\eqref{eq:red_models_final}) with their corresponding reference values from the database while also highlighting the values of $\varepsilon_{\mathrm{pred}}$.
All models exhibit $\varepsilon_{\mathrm{pred}}$ values consistently below $0.18$. 
Specifically, these errors range from $\varepsilon_{\mathrm{pred}} \approx 0.109$ for $\hat{\rho}= 0.7$ to $\varepsilon_{\mathrm{pred}} = 0.179$ for $\hat{\rho} = 0.3$. 
For context, considering the size of our validation datasets, this performance is superior to the existing results for reduced models for ETG turbulence in tokamak pedestals; for example, \citet{FMJ24} reported a minimum error $\varepsilon_{\mathrm{pred}} \approx 0.2$ for $40$ beyond-training predictions, while~\citet{Hatch_2024} achieved $\varepsilon_{\mathrm{pred}} \approx 0.15$ on reproducing their training dataset.

We also computed $95\%$ prediction intervals using bootstrapping, which, as shown in Figure~\ref{fig:red_model_pred_x0_0_2_to_0_8}, tend to narrow for the final training set sizes obtained from the active learning procedure. 
For a more comprehensive assessment of model performance, we instead show histograms of the relative prediction errors in the right panels of Figures~\ref{fig:red_model_pred_x0_0_2}--\ref{fig:red_model_pred_x0_0_8}, providing insight beyond pointwise predictions. 
These histograms indicate that a substantial portion of the relative errors fall below $20\%$, the majority remain under $50\%$, and only a relatively small number of outliers exceed this range.

\section{Extension of reduced models to additional radial positions}\label{sec:results_three_further_loc}

To provide a more complete assessment, we extend the presented analysis by using the fitted coefficients from Eq.~\eqref{eq:red_models_final} to construct generalized models that predict transport beyond the original seven radial locations.
Specifically, the fitted coefficients are used to derive regression-based expressions for the parameters of the target scaling law~\eqref{eq:target_model} as explicit functions of $\hat{\rho}$ (see Appendix~\ref{sec_app:coeff_formulas} for details).

Evaluating the coefficient formulae~\eqref{eq:curve_fit_coefficients} at the three additional radial positions, $\hat{\rho} \in \{0.1, 0.55, 0.75\}$, yields:
\begin{subequations}\label{eq:red_models_x0_0_1_0_55_0_75}
    \begin{align}
        \hat{\rho} = 0.1: \quad Q_e/Q_{\mathrm{GB}} & = 0.33 \, \omega_{T_e}^{2.83} \left(\eta_e - 1\right)^{1.06} \uptau^{-1.75} \label{eq:red_model_x0_0_1} \\
        \hat{\rho} = 0.55: \quad Q_e/Q_{\mathrm{GB}} & = 0.48 \, \omega_{T_e}^{2.34} \left(\eta_e - 1\right)^{1.13} \uptau^{-1.41} \label{eq:red_model_x0_0_55} \\
        \hat{\rho} = 0.75: \quad Q_e/Q_{\mathrm{GB}} & = 0.87 \, \omega_{T_e}^{1.79} \left(\eta_e - 1\right)^{1.17} \uptau^{-1.26}\,. \label{eq:red_model_x0_0_75}
    \end{align}
\end{subequations}

The coefficients of these generalized models are consistent with those in Eq.~\eqref{eq:red_models_final} for the original seven $\hat{\rho}$. 
Figure~\ref{fig:red_model_pred_x0_0_1_and_0_55_and_0_75} plots the predictions from the new models~\eqref{eq:red_models_x0_0_1_0_55_0_75} at the three additional positions. 
All models exhibit good agreement with the reference data, yielding deterministic errors of $0.168$, $0.167$, and $0.082$, respectively.
Moreover, the histograms of relative errors, shown in the right subplots, indicate that the majority of errors are below $20\%$ for $\hat{\rho} \in \{0.1, 0.75\}$ and below $40\%$ for $\hat{\rho} = 0.55$.
A relatively low number of outliers is observed at $\hat{\rho} = 0.1$.
Collectively, these results demonstrate that reduced models with coefficients obtained from the regression formulae~\eqref{eq:curve_fit_coefficients} retain predictive capability beyond the training radial locations. 
In particular, both interpolation and moderate extrapolation in $\hat{\rho}$ achieve predictive accuracy comparable to that at the original seven $\hat{\rho}$. 

\section{Summary and discussion}\label{sec:conclusions}
This paper investigated machine-learning-based reduced models for electron-temperature-gradient-driven (ETG) turbulence in the Wendelstein 7-X stellarator. 
Our approach, combining regression with an iterative active learning procedure, efficiently constructed reduced models depending on three key plasma parameters: the normalized electron temperature radial gradient ($\omega_{T_e}$), the ratio of normalized electron temperature and density radial gradients ($\eta_e$), and the electron-to-ion temperature ratio ($\uptau$). 
To our knowledge, this is the first study to derive and numerically validate ETG reduced models for a stellarator configuration.

We first constructed reduced models at seven radial locations, $\hat{\rho} \in \{0.2, 0.3, 0.4, 0.5, 0.6, 0.7, 0.8\}$, for which the active-learning procedure required training sets containing $104$--$190$ samples. 
Validation on more than $393$ out-of-sample points per $\hat{\rho}$ demonstrated that the models achieve good predictive accuracy. 
To extend the models beyond the original radial locations, we derived regression-based parameterizations for the coefficients appearing in the target scaling law. 
Numerical validation at three additional radial positions, $\hat{\rho} \in \{0.1, 0.55, 0.75\}$, including both interpolation and moderate extrapolation cases, showed that the coefficient regression models retained predictive accuracy in both regimes.

As the proposed methodology is primarily data-driven and numerically motivated, some limitations in applicability and physical consistency are expected. 
Although the models reproduce the expected leading-order ETG dependencies, further improvements may be achieved by incorporating additional physics into the model formulation. 
Future work will investigate extensions of the scaling law to account for geometry effects, $\eta_e$ threshold dependencies on additional plasma parameters, higher-order cross-coupling effects, and other mechanisms not captured in the current formulation.
Building on the present results, we also plan to extend the methodology to ion-scale transport quantities. 
In particular, reduced models for the ion-scale electron heat flux ($Q_{e,i}/Q_{\mathrm{GB}}$), ion heat flux ($Q_{i,i}/Q_{\mathrm{GB}}$), and particle flux ($\Gamma/\Gamma_{\mathrm{GB}}$) will be developed and validated in future work.

\begin{acknowledgments}
I.-G.~F. was supported in part by NSF Grant DMS-2436357.
This work has been carried out within the framework of the EUROfusion Consortium, funded by the European Union via the Euratom Research and Training Programme (Grant Agreement No 101052200 -- EURO-fusion) and by the Spanish Ministry of Science, Innovation and Universities under grant PID2021-125607NB-I00. 
We acknowledge the EuroHPC Joint Undertaking for awarding access to the EuroHPC supercomputer LUMI, hosted by CSC (Finland) and the LUMI consortium through a EuroHPC Extreme Access call. 
The simulations were completed under project grant number 465000821. 
Views and opinions expressed are however those of the author(s) only and do not necessarily reflect those of the European Union or the European Commission. 
Neither the European Union nor the European Commission can be held responsible for them. Numerical simulations were performed with the the Frontera supercomputer at the Texas Advanced Computing Center, USA, the Raven HPC system at the Max Planck Computing and Data Facility, Germany, the Marconi 100 \& Leonardo Fusion supercomputer at CINECA, Italy, and the LUMI supercomputer at the CSC data center, Finland.
\end{acknowledgments}

\appendix

\section{Sensitivity of the active learning approach to the choice of base models and testing sets} \label{sec_app:active_learning_sensitivity}
As discussed in the main text, three key components directly shape the modeling strategy adopted in this work: 
(i) the chosen model form~\eqref{eq:target_model}, 
(ii) the active learning procedure used for model construction, and 
(iii) the datasets on which the models are trained. 
The assumed model form and the available data largely determine the attainable predictive accuracy and robustness.
In the present work, we rely on a database generated in prior studies, and the active learning strategy is therefore used to identify the most informative training samples from this pre-existing pool of simulations. 
Although the available database is substantially larger than those used in earlier ETG reduced modeling studies, such as~\citet{Ha22,Hatch_2024} and~\citet{FMJ24}, which relied on only $\mathcal{O}(10)$ samples, it was not originally designed for surrogate model development. 
As a result, some parameters span a narrower range than that observed experimentally. 
For example, at $\hat{\rho}=0.5$, experimental discharges can reach values of $\uptau \sim 2.5$, whereas the database extends only to $1.94$.~\cite{Wappl2025}
Similarly, at $\hat{\rho}=0.7$, experiments reach values of $\uptau \sim 1.8$, while the database extends only to $1.37$.~\cite{Wappl2025} 
Consequently, regressions performed exclusively on the available database remain constrained by the parameter variation represented in the sampled data. 
While such fits are directly tied to the simulation database itself, they may still provide an incomplete characterization of parameter sensitivities across the experimentally relevant operating regime.
Within these constraints, the active learning framework provides a systematic mechanism for constructing informative training subsets relative to the chosen acquisition criterion and model ansatz.

Two critical questions arise. 
First, how sensitive is the approach to the choice of the testing set, which plays a central role in the active learning procedure? 
This question was addressed in the main text (see Figs.~\ref{fig:coeff_ev_x0_0_2_and_0_3}–\ref{fig:coeff_ev_x0_0_8} and the corresponding discussion). 
We showed that, for thresholds $\mathrm{max\_dev} = \mathrm{max\_rel\_err} = 0.10$ and $M = 10{,}000$ bootstrap samples, the active learning approach is not overly sensitive to the particular choice of testing dataset.
Second, how does the use of sparse grid data influence the construction of the initial models in the active-learning-based refinement process? 
An alternative strategy would be to initialize the model using a subset of the available database. 
However, the effectiveness of such an approach depends critically on the richness and representativeness of the database. 

In the following, we illustrate the influence of the initial sparse grid set at four radial locations, namely $\hat{\rho} \in \{0.3, 0.6, 0.7, 0.8\}$.
In all experiments, the testing set is identical to that used to obtain the fitted models reported in the main text. 
The base model is trained using a subset of the same size as the corresponding sparse grid, extracted pseudo-randomly from the remainder of the database. 
Consequently, the validation sets are slightly smaller by the number of points used to construct the base models. 
For simplicity, we present results for three independent fits using differently sampled subsets to construct the base models. 

We first consider $\hat{\rho} = 0.3$, where the fitted scaling law~\eqref{eq:red_model_x0_0_3} yielded the largest prediction error. 
The fitted models and corresponding prediction errors are:
\begin{equation*}
    Q_e/Q_{\mathrm{GB}} = 0.23 \, \omega_{T_e}^{2.38} \left(\eta_e - 1\right)^{1.31} \uptau^{-2.01},
    \quad \varepsilon_{\mathrm{pred}} = 0.182,
\end{equation*}
\begin{equation*}
    Q_e/Q_{\mathrm{GB}} = 0.23 \, \omega_{T_e}^{2.42} \left(\eta_e - 1\right)^{1.28} \uptau^{-1.99}, 
    \quad \varepsilon_{\mathrm{pred}} = 0.179,
\end{equation*}
\begin{equation*}
    Q_e/Q_{\mathrm{GB}} = 0.23 \, \omega_{T_e}^{2.37} \left(\eta_e - 1\right)^{1.32} \uptau^{-2.00}, 
    \quad \varepsilon_{\mathrm{pred}} = 0.182.
\end{equation*}
The size of the training set varied between $178$ and $182$. 
Although small variations in the fitted parameters are observed, they remain minor and the overall predictive performance is essentially unchanged.

Next, we consider $\hat{\rho} = 0.6$, for which we obtain:
\begin{equation*}
    Q_e/Q_{\mathrm{GB}} = 0.47 \, \omega_{T_e}^{1.99} \left(\eta_e - 1\right)^{1.19} \uptau^{-0.70}, 
    \quad \varepsilon_{\mathrm{pred}} = 0.172,
\end{equation*}
\begin{equation*}
    Q_e/Q_{\mathrm{GB}} = 0.50 \, \omega_{T_e}^{1.96} \left(\eta_e - 1\right)^{1.19} \uptau^{-0.79}, 
    \quad \varepsilon_{\mathrm{pred}} = 0.169,
\end{equation*}
\begin{equation*}
    Q_e/Q_{\mathrm{GB}} = 0.55 \, \omega_{T_e}^{1.84} \left(\eta_e - 1\right)^{1.19} \uptau^{-0.59}, 
    \quad \varepsilon_{\mathrm{pred}} = 0.172.
\end{equation*}
The training set size varied between $143$ and $146$. 
The fitted powers of $\omega_{T_e}$ and $\left(\eta_e - 1\right)$ remain broadly consistent with those reported in Eq.~\eqref{eq:red_model_x0_0_6}, while the largest variations occur in the exponent of $\uptau$. 
The validation error remains consistent with that of the model reported in the manuscript.

The variations in the fitted exponent of $\uptau$ are most pronounced at $\hat{\rho} = 0.7$, where we obtain:
\begin{equation*}
    Q_e/Q_{\mathrm{GB}} = 1.12 \, \omega_{T_e}^{1.59} \left(\eta_e - 1\right)^{0.92} \uptau^{0.14}, 
    \quad \varepsilon_{\mathrm{pred}} = 0.100,
\end{equation*}
\begin{equation*}
    Q_e/Q_{\mathrm{GB}} = 1.20 \, \omega_{T_e}^{1.55} \left(\eta_e - 1\right)^{0.92} \uptau^{0.09}, 
    \quad \varepsilon_{\mathrm{pred}} = 0.092,
\end{equation*}
\begin{equation*}
    Q_e/Q_{\mathrm{GB}} = 1.13 \, \omega_{T_e}^{1.57} \left(\eta_e - 1\right)^{0.95} \uptau^{-0.05}, 
    \quad \varepsilon_{\mathrm{pred}} = 0.099.
\end{equation*}
The training size varied between $65$ and $71$. 
The fitted values of $p_3$ are inconsistent with those obtained at other radial locations and yield physically implausible exponents close to zero or even positive. 
A value of $p_3$ consistent with other radial locations (i.e., $p_3 \sim -1$) was obtained only when the sparse grid data were included in the testing set during the active learning procedure, corresponding to our original approach.

Finally, the three independent fits without sparse grid initialization at $\hat{\rho} = 0.8$ yield
\begin{equation*}
    Q_e/Q_{\mathrm{GB}} = 1.23 \, \omega_{T_e}^{1.44} \left(\eta_e - 1\right)^{1.16} \uptau^{-2.57}, 
    \quad \varepsilon_{\mathrm{pred}} = 0.109,
\end{equation*}
\begin{equation*}
    Q_e/Q_{\mathrm{GB}} = 1.19 \, \omega_{T_e}^{1.45} \left(\eta_e - 1\right)^{1.18} \uptau^{-2.42}, 
    \quad \varepsilon_{\mathrm{pred}} = 0.106,
\end{equation*}
\begin{equation*}
    Q_e/Q_{\mathrm{GB}} = 1.21 \, \omega_{T_e}^{1.44} \left(\eta_e - 1\right)^{1.17} \uptau^{-2.36}, 
    \quad \varepsilon_{\mathrm{pred}} = 0.108.
\end{equation*}
The training size varied between $98$ and $104$.

In all experiments, the validation errors of these models remain consistent with the corresponding errors reported in Table~\ref{tab:training_and_validation_eps}. 
The relatively large shifts in $p_3$ across different $\hat{\rho}$, from approximately $-2.00$ at $\hat{\rho} = 0.3$, to approximately $-0.7$ at $\hat{\rho} = 0.6$, then approximately $0.0$ at $\hat{\rho} = 0.7$, and finally approximately $-2.40$ at $\hat{\rho} = 0.8$, reflects the structure and coverage of the available database more than indicating physically consistent trends. 
Using sparse grid data to construct the base models leads to more consistent values of $p_3$ across radial locations. 
As noted above, the resulting fit is influenced by parameter coverage of the available training data.

\section{Regression formulae for the scaling law coefficients} \label{sec_app:coeff_formulas}
The fitted model coefficients in Eq.~\eqref{eq:red_models_final} indicate nonlinear trends for $c_0$ and $p_1$, and approximately linear trends for $p_2$ and $p_3$ with respect to the radial location, $\hat{\rho}$. 
To construct robust coefficient regression models, reduce the risk of overfitting, and account for the nonmonotonic trends in the fitted coefficients, we select quadratic polynomials for $c_0$ and $p_1$, and linear functions for $p_2$ and $p_3$.
These choices are made purely for mathematical convenience and are not intended to imply any physical interpretation. 
To further enhance the robustness of the fits, outliers were removed from the curve-fitting procedure.
The resulting expressions for the model coefficients are: 
\begin{subequations}\label{eq:curve_fit_coefficients}
    \begin{align}
        c_0(\hat{\rho}) &= 2.47\,\hat{\rho}^2 - 1.27\,\hat{\rho} + 0.43, \label{eq:curve_fit_c0} \\
        p_1(\hat{\rho}) &= -2.53\,\hat{\rho}^2 + 0.55\,\hat{\rho} + 2.80, \label{eq:curve_fit_p1} \\
        p_2(\hat{\rho}) &= 0.17\,\hat{\rho} + 1.04, \label{eq:curve_fit_p2} \\
        p_3(\hat{\rho}) &= 0.76\,\hat{\rho} - 1.83. \label{eq:curve_fit_p3}
    \end{align}
\end{subequations}

\section{Investigation into a radially independent reduced model} \label{sec_app:generic_model}

For completeness, we also considered constructing a single generic scaling ansatz~\eqref{eq:target_model} with coefficients independent of the radial location. 
To this end, we applied the active learning procedure described in Sec.~\ref{sec:ML_based_reduced_modeling} to the aggregated data from all available radial positions. 
Because the sparse-grid parameter bounds are identical for all $\hat{\rho}$, the initial base model was constructed using data from the central radial location, $\hat{\rho}=0.5$. 
The testing set during the active learning procedure consisted of $\lfloor N_{\mathrm{DB}}/20 \rfloor$ randomly selected database inputs from each radial location. 
The iterative refinement procedure was terminated once both $\mathrm{max\_dev}$ and $\mathrm{max\_rel\_err}$ fell below $10\%$. 
This configuration was chosen after systematic experimentation with different testing-set sizes and threshold values. 
The final training set contains $N_{\mathrm{train}} = 225$ input--output pairs, and the validation set comprises $N_{\mathrm{pred}} = 5{,}011$ data points.

The resulting generic scaling law reads:
\begin{equation}
\label{eq:generic_model}
    Q_e/Q_{\mathrm{GB}}
    =
    0.28 \,
    \omega_{T_e}^{2.23}
    \left(\eta_e - 1\right)^{1.39}
    \uptau^{-2.21}.
\end{equation}
The fitted coefficients are broadly consistent with those obtained for the seven radially resolved models, with a modest increase in the exponent of $(\eta_e - 1)$ and a slight decrease in the exponent of $\uptau$.

The deterministic training error is $\varepsilon_{\mathrm{train}} = 0.188$, comparable to the training errors of the individual radial models. 
The generic scaling law achieves a validation error of $\varepsilon_{\mathrm{pred}} = 0.201$ against the aggregated validation dataset from all seven radial locations, $\hat{\rho} \in \{0.2,0.3,0.4,0.5,0.6,0.7,0.8\}$, indicating reasonable predictive capability over a broad parameter space. 
However, the generic model fails to provide consistently accurate predictions at the three additional radial locations, $\hat{\rho} \in \{0.1,0.55,0.75\}$. 
While predictions at $\hat{\rho} = 0.1$ are consistent with the regression-fitted model~\eqref{eq:red_model_x0_0_1} ($\varepsilon_{\mathrm{pred}} = 0.157$), the generic model systematically overpredicts the flux at $\hat{\rho} = 0.55$ and $0.75$, yielding errors of $\varepsilon_{\mathrm{pred}} = 0.30$ and $0.21$, respectively. 

These results suggest that a purely radially independent scaling law of the form~\eqref{eq:target_model} is limited by one or more factors not captured by our small set of input parameters, such as variations in ETG-driven transport across flux surfaces, local magnetic geometry, etc. 
As it stands, this generalized model is not recommended to be used as a predictive tool and is presented solely to show the limitations of the chosen functional form and the available dataset.
A more systematic investigation of globally consistent yet expressive transport models is left for future work.

\bibliography{bibliography}

\end{document}